\documentclass[a4paper]{article}

\usepackage{amsmath,amsfonts}
\usepackage[ruled]{algorithm2e}

\usepackage{subfigure}
\usepackage{enumerate}
\usepackage{wrapfig}
\usepackage{hyperref}
\usepackage{pgfplots}
\usepackage{pgfplotstable}
\usepackage{tikz}
\usetikzlibrary{positioning}
\usetikzlibrary{trees}
\usetikzlibrary{backgrounds}
\usetikzlibrary{arrows,shapes}
\usetikzlibrary{mindmap}
\usetikzlibrary{chains}
\usetikzlibrary{calc}
\usetikzlibrary{matrix}
\usetikzlibrary{decorations.pathmorphing,patterns,shadows,decorations.markings}
\usepgfplotslibrary{fillbetween}
\usepgfplotslibrary{colormaps}

\tikzset{
  set arrow inside/.code={\pgfqkeys{/tikz/arrow inside}{#1}},
  set arrow inside={end/.initial=>, opt/.initial=},
  /pgf/decoration/Mark/.style={
    mark/.expanded=at position #1 with
    {
      \noexpand\arrow[\pgfkeysvalueof{/tikz/arrow inside/opt}]{\pgfkeysvalueof{/tikz/arrow inside/end}}
    }
  },
  arrow inside/.style 2 args={
    set arrow inside={#1},
    postaction={
      decorate,decoration={
        markings,Mark/.list={#2}
      }
    }
  },
}

\SetAlFnt{\small}
\SetAlCapFnt{\small}
\SetAlCapNameFnt{\small}
\SetAlCapHSkip{0pt}
\IncMargin{-\parindent}

\newcommand{\ve}[1]{\mathbf{#1}}
\newcommand{\avg}[1]{\{\!\{#1\}\!\}}
\newcommand{\jump}[1]{[\![#1]\!]}
\newcommand{\trans}{\mathrm{T}}

\newtheorem{Remark}{Remark}
\newtheorem{Requirement}{Design Choice}

\makeatletter
\definecolor{gnuplot@orange}{RGB}{229,158,0}
\definecolor{gnuplot@purple}{RGB}{148,0,212}
\definecolor{gnuplot@lightblue}{RGB}{87,181,232}
\definecolor{gnuplot@green}{RGB}{0,158,75}
\definecolor{gnuplot@darkblue}{RGB}{0,115,179}
\definecolor{gnuplot@yellow}{RGB}{240,227,66}
\pgfplotscreateplotcyclelist{colorGPL}{%
gnuplot@darkblue,every mark/.append style={fill=gnuplot@darkblue!80!black},mark=o\\%
red!80!white,every mark/.append style={fill=red!80!black},mark=square\\%
gnuplot@green,every mark/.append style={fill=gnuplot@green!50!black},mark=otimes*\\%
gnuplot@orange,every mark/.append style={fill=gnuplot@orange!80!black,mark size=2.5pt},mark=x\\%
gnuplot@purple,mark=diamond\\%
black,densely dashed,every mark/.append style={solid,fill=gnuplot@darkblue!80!black},mark=square*\\%
gnuplot@lightblue,densely dashed,every mark/.append style={solid,fill=gnuplot@lightblue!30!black},mark=triangle*\\%
red!80!white,densely dashed,every mark/.append style={solid,fill=red!40!white},mark=oplus*\\%
}
\newcounter{datastructure}
\newenvironment{datastructure}[1][htb]
  {
   \let\c@algocf\c@datastructure
   \begin{algorithm}[#1]%
  }{\end{algorithm}}
\makeatother

\pgfplotsset{compat=1.9}

\input{tables.tex}

\newcommand{\TheTitle}{Fast matrix-free evaluation of discontinuous Galerkin finite element operators}

\title{{\TheTitle}\thanks{  This work was supported by the German Research Foundation (DFG) under the
  project ``High-order discontinuous Galerkin for the exa-scale'' (ExaDG)
  within the priority program ``Software for Exascale Computing'' (SPPEXA),
  grant agreement no.~KO5206/1-1 and KR4661/2-1. The authors acknowledge the
  support given by the Bayerische Kompetenznetzwerk f\"ur
  Technisch-Wissenschaftliches Hoch- und H\"ochstleistungsrechnen (KONWIHR) in
  the framework of the project \emph{Matrix-free GPU kernels for complex
    applications in fluid dynamics}. The authors gratefully acknowledge the
  Gauss Centre for Supercomputing e.V.~(\texttt{www.gauss-centre.eu}) for
  funding this project by providing computing time on the GCS Supercomputer
  SuperMUC at Leibniz Supercomputing Centre (LRZ, \texttt{www.lrz.de}) through
  project id pr83te.
}}

\author{
  Martin Kronbichler\thanks{Institute for Computational Mechanics, Technical University of Munich, Boltzmannstr.~15, 85748 Garching b.~M\"unchen, Germany
    (\texttt{kronbichler@lnm.mw.tum.de}).}
  \and
  Katharina Kormann\thanks{
  Max Planck Institute for Plasma Physics, Boltzmannstr.~2, 85748 Garching, Germany, and Zentrum Mathematik, Technical University of Munich, Boltzmannstr.~3, 85748 Garching, Germany (\texttt{katharina.kormann@ipp.mpg.de}).}
}

\usepackage{amsopn}

\begin{document}

\maketitle

\begin{abstract}
We present an algorithmic framework for matrix-free evaluation of discontinuous Galerkin finite element operators based on sum factorization on quadrilateral and hexahedral meshes. We identify a set of kernels for fast quadrature on cells and faces targeting a wide class of weak forms originating from linear and nonlinear partial differential equations. Different algorithms and data structures for the implementation of operator evaluation are compared in an in-depth performance analysis. The sum factorization kernels are optimized by vectorization over several cells and faces and an even-odd decomposition of the one-dimensional compute kernels. In isolation our implementation then reaches up to 60\% of arithmetic peak on Intel Haswell and Broadwell processors and up to 50\% of arithmetic peak on Intel Knights Landing. The full operator evaluation reaches only about half that throughput due to memory bandwidth limitations from loading the input and output vectors, MPI ghost exchange, as well as handling variable coefficients and the geometry. Our performance analysis shows that the results are often within 10\% of the available memory bandwidth for the proposed implementation, with the exception of the Cartesian mesh case where the cost of gather operations and MPI communication are more substantial.
\end{abstract}


\section{Introduction}

The discontinuous Galerkin (DG) method gained a lot of momentum in a wide
range of applications in the last two decades. The method combines the
favorable features of the numerical fluxes in finite
volume methods, often called Riemann solvers, with the high-order capabilities of finite elements based on
polynomial bases. This construction allows for both high-order convergence
rates on complicated computational domains as well as robustness in transport-dominated
problems. DG methods have been identified as one of the most promising schemes
for next-generation solvers in fluid dynamics and wave propagation problems and have also been
introduced to a large number of other problems with mixed first and second order
derivatives.

There is a large body of literature on implementing DG schemes and performance tuning for
particular equations, especially for GPUs, see e.g.
\cite{kwbh09,Modave16,Abdi17} and references therein. Classical matrix-based implementations for explicit time
integration and various optimizations for triangles and tetrahedra have reached a high level of
maturity \cite{hw08}. The focus of this work is on optimized shape value
interpolation and derivative kernels for DG methods on quadrilaterals and
hexahedra with moderate polynomial degrees in the range between 2 and 10 in
the context of general meshes and variable coefficients. There exist
special techniques for Cartesian meshes where the final cell matrix has tensor
product structure \cite{lynch64,huismann17}, but in the general case of more complex geometries and variable coefficient
the final stencil cannot be separated into tensor products of 1D stencils. In that case, the fastest implementation option
is usually the evaluation of integrals on the fly by fast sum
factorization techniques \cite{Deville02,KS05,Kopriva09} that have their origin
in spectral elements \cite{Orszag80,Patera84}. These methods have an operator
evaluation cost of $\mathcal O(dk)$ per degree of freedom in the spatial dimension $d$ for $k$ polynomials per direction, i.e., polynomial degree $p=k-1$, in contrast to costs of
$\mathcal O(k^d)$ in matrix-based variants. While originally used for higher degrees, recent work
\cite{brown10,Kronbichler12,Kronbichler16b} has shown that these techniques are
up to an order of magnitude faster than sparse matrix kernels already for medium
polynomial degrees of $p=3$ or $p=4$, with increasing gaps at higher orders.

Tensor product evaluation has been a very active research area with implementations available in the generic finite element software packages
DUNE \cite{exadune-14,Bastian2016}, Firedrake
\cite{rathgeber16,mcrae16,luporini17}, Loopy \cite{kloeckner14}, mfem
\cite{mfem-web-page}, Nek5000 \cite{nek5000-web-page}, Nektar++
\cite{cantwell15}, or NGSolve \cite{schoeberl14} as well as application codes
such as the compressible flow solver framework Flexi
\cite{Hindenlang12}, SPECFEM3D \cite{SPECFEM3D} or pTatin3D
\cite{maybrown14}. Despite the wide availability of software, including code
generators and domain-specific languages in Firedrake and Loopy, we believe that the analysis of high performance computing aspects and the expected performance envelopes of operator evaluation---independent
of the user interfaces---are still missing. This work fills this gap by
presenting an extensive analysis of the DG matrix-free operator evaluation
with focus on the choice of data layouts and loop structures with their
respective impact on performance for architectures with a deep memory hierarchy, supported by careful
numerical experiments and results from hardware performance counters
that quantify the effect of optimizations. Our focus is on three specific topics:
\begin{itemize}
\item efficient local kernels for shape value interpolation and derivatives
  with code vectorization over several elements,
\item the storage of degrees of freedom with this vectorization scheme, and
\item aspects of efficient MPI parallelization.
\end{itemize}
We concentrate on high-quality
kernels for medium polynomial degrees $2\leq p\leq 10$, but we show
results also beyond this number to illustrate that there are no sudden
breakdowns in performance. The proposed algorithms aim for reaching high single-node performance,
but they also directly apply to the
massively parallel context using the scheme presented in \cite{bbhk11} as has
been demonstrated on up to 147,456 cores in
\cite{Kronbichler16b,Krank17}, given that the communication in operator
evaluation is only between nearest neighbors and thus naturally scalable to
large processor counts. To the best of the authors' knowledge, no similarly detailed
analysis of DG algorithms for cache-based hardware has been previously
presented in the literature. Selected well-performing options of the presented
work are included in a comprehensive kernel collection of the
open-source finite element library \texttt{deal.II} \cite{dealII85} and are
accessible to generic applications within a toolbox providing other advanced features such
as multigrid solvers and mesh adaptivity with hanging nodes. The developments
in this work interoperate with the continuous finite element implementations from
\cite{Kronbichler12} using the same optimized code paths in the relevant
algorithms and have already been used in a number of very challenging application
codes in fluid dynamics \cite{Kronbichler16a,Krank17} and wave propagation
\cite{Kormann16,Kronbichler16}. Note that operator evaluation is the central
algorithmic part not only in explicit time integration but also for some
implicit solvers such as multigrid methods with selected smoothers
\cite{Carr16,Kronbichler16b}.

This article is structured as follows. Sec.~\ref{sec:algorithm} introduces the
mathematics of the underlying partial differential equations, the DG
discretization, and the matrix-free implementation based on fast
integration. Sec.~\ref{sec:kernel_local} concentrates on the compute part of
the algorithm, i.e., the tensor product kernels. The access to the solution
vectors for cell and face integrals are analyzed in
Sec.~\ref{sec:vector_access} alongside with the question of efficient ghost
data exchange with MPI. Sec.~\ref{sec:geometry} discusses the options for
how to apply the geometry
and Sec.~\ref{sec:conclusions} concludes this
work.

\section{DG algorithm}\label{sec:algorithm}

We assume a triangulation $\mathcal T_h$ on the computational domain $\Omega$
that consists of a set of elements $\Omega_e$. We denote the faces
$\mathcal F_h$ as the set of all intersections
$\bar{\Omega}_{e^-}\cap \bar{\Omega}_{e^+}$, i.e., the edges of the cells in 2D
and the surfaces of the cells in 3D, with the subset
$\mathcal F_h^{\mathrm{i}}$ denoting the interior faces between two cells
$\Omega_{e^-}$ and $\Omega_{e^+}$ with solution $u_h^-$ and $u_h^+$, respectively, and
the set of boundary faces $\mathcal F_h^{\mathrm{b}}$ where only the
solution field $u_h^-$ is present. The vectors $\ve n^{-}=-\ve n^{+}$ denote
the outer unit normal vectors on the two sides of a face. We also write $\ve n$
instead of $\ve n^{-}$ for the normal vector associated to the cell under
consideration $\Omega_e$.

The quantity $\avg{u} = \frac{1}{2}(u_h^-+u_h^+)$ denotes the average of the
values on the two sides of a face and the jump is written as
$\jump{u_h} = u_h^- \ve n^- + u_h^+ \ve n^+ = (u_h^--u_h^+) \ve n^-$. At domain
boundaries, suitable definitions for the exterior solution $u_h^+$ in terms
of the boundary conditions and the inner solution $u_h^-$ are used, e.g.~the
mirror principle $u_h^+=-u_h^-+2g$ in case of Dirichlet conditions
\cite{hw08}. Inhomogeneous Dirichlet data add contributions to the right hand
side vectors in linear systems.

To exemplify our algorithms and implementations, we consider two prototype
discontinuous Galerkin discretizations to stationary problems: The DG
discretization of the stationary advection equation with local Lax--Friedrichs
(upwind) flux \cite{hw08} is characteristic for first-order hyperbolic PDEs
and reads
\begin{equation}\label{eq:advection}
  -(\nabla v_h,\ve c u_h)_{\Omega_e} + \left< v_h \ve n,\avg{\ve c u_h} + \frac{|\ve c \cdot \ve n|}{2} \jump{u_h}\right>_{\partial \Omega_e}  =(v_h, f)_{\Omega_e},
\end{equation}
where $\ve c = \ve c(\ve x)$ denotes the direction of transport and $f$ is a
forcing function. The bilinear form
$(a,b)_{\Omega_e} = \int_{\Omega_e}ab\,\mathrm{d}\ve x$ denotes volume
integrals and
$\left<a,b\right>_{\partial \Omega_e} = \int_{\partial
  \Omega_e}ab\,\mathrm{d}\ve s$
boundary integrals.

The symmetric interior penalty discretization of the Laplacian \cite{arnold02} is given by the weak form
\begin{equation}\label{eq:poisson_sip}
  (\nabla v_h,\nabla u_h)_{\Omega_e} - \left<\nabla v_h,\frac{1}{2}\jump{u_h}\right>_{\partial \Omega_e} - \left<v_h \ve n,\avg{\nabla u_h}\right>_{\partial \Omega_e} + \left<v_h \ve n,\tau \jump{u_h}\right>_{\partial \Omega_e}
  =(v_h, f)_{\Omega_e},
\end{equation}
on element $\Omega_e$, containing the cell integral, the adjoint consistency
term, the primal consistency term, and an interior penalty term with factor $\tau$ sufficiently large to render the
discretization coercive.

Since the operators on the left
hand sides of equations \eqref{eq:advection} and \eqref{eq:poisson_sip} are linear, they correspond to a matrix-vector product
taking a vector of coefficient values $\ve u = [u_i]$ associated with the
solution field $u_h$ and returning the integrals $\ve y = [y_i]$ tested by test functions
$v_h=\varphi_i$,
\begin{equation}\label{eq:matvec}
\ve y = A \ve u.
\end{equation}
Since our methods are based on numerical integration rather than the final
cell matrices, the techniques extend straight-forwardly to nonlinear
equations.

On each element $\Omega_e$, we assume the solution to be given by an expansion
$u_h^{(e)} = \sum_{i=1}^{N} \varphi_i(\ve x) u_i^{(e)}$, where $u_i^{(e)}$ are
the coefficient values determined through the variational principle and
$\varphi_i(\ve x)$ are polynomial basis functions. For the definition of the
integrals, we transform the basis functions on the reference element
$\Omega_\text{unit}$ with coordinates $\boldsymbol \xi$ to the real element
coordinates $\ve x$ by a transformation $\hat{\ve x}^{(e)}$ as
$\ve x = \hat{\ve x} ^{(e)}(\boldsymbol \xi)$ with Jacobian
$\mathcal J^{(e)} = \frac{\mathrm d \hat{\ve x}^{(e)}}{\mathrm d \boldsymbol
  \xi} = \nabla_{\boldsymbol \xi} \hat{\ve x}^{(e)}$. Thus, the derivative can
be expressed as
$\nabla_{\ve x} \varphi(\boldsymbol{\xi}(\ve x)) = \left(\mathcal
  J^{(e)}\right)^{-\trans} \nabla_{\boldsymbol \xi} \varphi(\boldsymbol \xi)$.
In this work, we assume quadrilateral elements in 2D or hexahedral elements in
3D with shape functions defined through the tensor product of one-dimensional
functions
$\varphi_i(\xi_1,\xi_2,\xi_3) = \varphi_{i_1}^\text{1D}(\xi_1)
\varphi_{i_2}^\text{1D}(\xi_2) \varphi_{i_3}^\text{1D}(\xi_3)$ with the
respective multi-index $(i_1,i_2,i_3)$ associated to the index~$i$.  The
integrals in equations \eqref{eq:advection} and \eqref{eq:poisson_sip} are
computed through numerical integration on a set of quadrature points
$\boldsymbol \xi_q = (\xi_{q_1},\xi_{q_2},\xi_{q_3})$ with associated weights
$w_q$ defined as the tensor product of 1D quadrature formulas. For example,
the cell term for advection in \eqref{eq:advection} is approximated by
\begin{equation}\label{eq:integral_advection}
  \begin{aligned}
    (\nabla \varphi_i, \ve c u_h)_{\Omega_e}&=\int_{\Omega_\text{unit}} \left(\mathcal J^{(e)}(\boldsymbol \xi)^{-\trans} \nabla _{\boldsymbol \xi} \varphi(\boldsymbol \xi) \right)\cdot \left(\ve c\left(\hat{\ve x}^{(e)}(\boldsymbol{\xi})\right) u_h^{(e)}(\boldsymbol \xi) \right) \mathrm{det}(\mathcal J^{(e)}(\boldsymbol \xi))\, \mathrm{d}\boldsymbol{\xi}\\
    & \approx \sum_{q=1}^{n_q} \left(\mathcal J^{(e)}(\boldsymbol \xi_q)^{-\trans}\nabla \varphi_i(\boldsymbol \xi_q)\right)\cdot \left(\ve c\left(\hat{\ve
          x}^{(e)}(\boldsymbol \xi_q)\right) u_h^{(e)}(\boldsymbol \xi_q)\right)
    \mathrm{det}(\mathcal J^{(e)}(\boldsymbol \xi_q)) w_q.
  \end{aligned}
\end{equation}

\subsection{Algorithm outline for discontinuous Galerkin finite element operator evaluation}

The matrix-free evaluation of the integrals representing the product
\eqref{eq:matvec} is implemented by a loop over all the
cells and faces appearing in the operators \eqref{eq:advection} or
\eqref{eq:poisson_sip}. The procedure for
the advection operator \eqref{eq:advection} is outlined in Algorithm
\ref{alg:prototype_dof_op}.

\begin{algorithm}
\caption{DG integration loop for the advection operator \eqref{eq:advection} with Dirichlet b.c.}
\label{alg:prototype_dof_op}
\begin{enumerate}[(i)]
  \setlength{\itemsep}{0pt}
  \setlength{\parsep}{0pt}
  \setlength{\parskip}{0pt}
\item {\label{alg:prototype_ghosts} {\tt update\_ghost\_values}: Import
    vector values of $\ve u$ from other MPI processes that are needed for
    computations of face integrals on the faces associated with the current
    process.}
\item {\label{alg:prototype_cell}
  loop over cells
  \begin{enumerate}
  \setlength{\itemsep}{0pt}
  \setlength{\parsep}{0pt}
  \setlength{\parskip}{0pt}
  \item \label{alg:cell_read} gather local vector values $u^{(e)}_i$
    on cell from global input vector $\ve u$
  \item \label{alg:cell_interpolate} interpolate local vector values
    $\ve u^{(e)}$ onto quadrature points,
    $u_h^{(e)}(\boldsymbol \xi_q) = \sum_i \varphi_i u^{(e)}_i$
  \item \label{alg:cell_geometry} for each quadrature index $j$, prepare
    integrand in each quadrature point by computing
    $\ve t_q = \mathcal J^{(e)}(\boldsymbol \xi_q)^{-1}\ve c\left(\hat{\ve
    x}^{(e)}(\boldsymbol \xi_q)\right) u_h^{(e)}(\boldsymbol \xi_q)
    \mathrm{det}(\mathcal J^{(e)}(\boldsymbol \xi_q)) w_q$
  \item \label{alg:cell_integrate} evaluate local integrals by quadrature
    $y^{(e)}_i = \left(\nabla \varphi_i,\ve c u_h^{(e)}\right)_{\Omega_e}
    \approx \sum_q \nabla \varphi_i(\boldsymbol \xi_q)\cdot \ve t_q$ for
    all test functions $i$
  \item \label{alg:cell_write} write the local
    contributions $y^{(e)}_i$ into the global result vector $\ve y$
     \end{enumerate}
}
\item {\label{alg:prototype_face}
  loop over interior faces
  \begin{enumerate}
  \setlength{\itemsep}{0pt}
  \setlength{\parsep}{0pt}
  \setlength{\parskip}{0pt}
  \item[{(a${}_-$)}] \label{alg:face_read1} gather local vector values $u^{-}_i$
     from global input vector $\ve u$ associated with interior cell $e^-$
  \item[{(b${}_-$)}] \label{alg:face_interpolate1} interpolate $u^-$ onto face quadrature
    points $u_h^{-}(\boldsymbol \xi_q)$
  \item[{(a${}_+$)}] \label{alg:face_read2}gather local vector values $u^{+}_i$
    from global input vector $\ve u$ associated with exterior cell $e^+$
  \item[{(b${}_+$)}] \label{alg:face_interpolate2}interpolate $u^+$ onto face quadrature
    points $u_h^{+}(\boldsymbol \xi_q)$
  \item[{(c)}] \label{alg:face_num_flux} for each quadrature index $j$, compute the
    numerical flux contribution
    $(\ve f^*\cdot \ve n^-)_q=\frac{1}{2}  \ve c\left(\hat{\ve
        x}^{(e)}(\boldsymbol \xi_q)\right) \cdot \ve n^{-} \left(u_h^{-}(\boldsymbol \xi_q) +
    u_h^{+}( \boldsymbol \xi_q)\right) + \frac{1}{2} \left|\ve
      c(\hat{\ve x}^{(e)}(\boldsymbol \xi_q))\cdot \ve n^{-}\right|\left(u_h^{-}(\boldsymbol
    \xi_q) - u_h^{+}( \boldsymbol \xi_q)\right)$ and multiply it by integration
    weight, $t_q = (\ve f^* \cdot \ve n^{-})_q h(\boldsymbol \xi_q) w_q$ with area element $h$ of face
  \item[{(d${}_-$)}] \label{alg:face_integrate1}evaluate local integrals by quadrature
    $y_i^{e^-} = (\varphi_i^-,\ve f^*\cdot \ve n^{-}) \approx \sum_q \varphi_i(\boldsymbol
      \xi_q) t_q$
  \item[{(e${}_-$)}] \label{alg:face_write1}add local contribution $y_i^{e^-}$ into the
    global result vector vector $\ve y$ associated with $e^-$
  \item[{(d${}_+$)}] \label{alg:face_integrate2}evaluate local integrals by quadrature
    $y_i^{e^+} = (\varphi_i^+,-\ve f^*\cdot \ve n^{-}) \approx \sum_q
    -\varphi_i(\boldsymbol \xi_q) t_q$ due to
    $\boldsymbol{n}^+=-\boldsymbol{n}^-$ and the conservativity of the numerical flux function
  \item[{(e${}_+$)}] \label{alg:face_write2}add local contribution $y_i^{e^+}$ into the
    global result vector vector $\ve y$ associated with $e^+$
  \end{enumerate}
}
\item {\label{alg:prototype_bound}
  loop over boundary faces
  \begin{enumerate}
  \setlength{\itemsep}{0pt}
  \setlength{\parsep}{0pt}
  \setlength{\parskip}{0pt}
  \item \label{alg:bound_read} gather local vector values $u^{-}_i$
    from cell $e^-$ from global input vector $\ve u$
  \item \label{alg:bound_interpolate} interpolate $u^-$ onto face quadrature
    points $u_h^{-}(\boldsymbol \xi_q)$
  \item \label{alg:bound_num_flux} for each quadrature index $j$, compute the
    numerical flux contribution
    $(\ve f^*\cdot \ve n^-)_q=\left| \ve c\left(\hat{\ve x}^{(e)}(\boldsymbol
        \xi_q)\right)\cdot \ve n^{-} \right|u_h^{-}(\boldsymbol \xi_q)$ and
    multiply by integration factor,
    $t_q = (\ve f^*\cdot \ve n^-)_q h(\boldsymbol \xi_q) w_q$
  \item \label{alg:bound_integrate}evaluate local integrals by quadrature
    $y_i^{e^-} = (\varphi_i^-,f^*) \approx \sum_q \varphi_i(\boldsymbol
      \xi_q) t_q$
  \item \label{alg:bound_write}add local contribution $y_i^{e^-}$ into the
    global result vector vector $\ve y$ associated with $e^-$
  \end{enumerate}
}
\item {\label{alg:prototype_compress} {\tt compress}: Export parts of the
    residuals that have been generated on the current MPI process to the
    owning process.}
\end{enumerate}
\end{algorithm}

Algorithm \ref{alg:prototype_dof_op} is split into three phases, addressing
the cell integrals, integrals for interior faces, and integrals for boundary
faces in separate steps. Each of these loops logically consists of five
components, which are (a) the extraction of the solution values pertaining to the
current cell(s) by a gather operation, (b) the evaluation of values or gradients
of the local solution in the quadrature points, (c) the operation on quadrature points including the application of the geometry, (d) the multiplication by the test functions and the summation in the numerical
quadrature, and finally (e) accumulating the local values into the respective entry of
the global vector by a scatter-add operation. Since each cell has
independent degrees of freedom in DG, the algorithm can skip zeroing the
result vector as typical in continuous finite elements \cite{Kronbichler12}
and instead set the integrals computed in the cell part (e), assuming that all
relevant cell integrals are done before the respective face integrals.

In Algorithm \ref{alg:prototype_dof_op}, the steps (b) and (d) interpolating the solution
from the coefficient values to quadrature points and quadrature loops are expensive in a naive implementation for higher
polynomial degrees because all degrees of freedom inside a cell can contribute
to the values on each quadrature point. For tensor product shape functions (or
truncated tensors) on tensor product quadrature formulas, highly efficient
schemes by sum factorization techniques as established in the spectral element
community are available \cite{Orszag80,Deville02,KS05,Kopriva09}.

Let us denote by $S_{i}$ the $k\times k$ matrix of values of all $k$ 1D shape functions of degree $p=k-1$ evaluated in $k$ quadrature points and by
$D_{i}$ the matrix of their derivatives along direction $i$, respectively. Further, denote by $\ve u^{(e)}$ the DG coefficients of the input vector. The evaluation of the $d$ components of the gradient $\nabla_{\boldsymbol{\xi}} u_h^{(e)}$ in
$d$-dimensional space (step (b) of a variant of Algorithm \ref{alg:prototype_dof_op} applied to the Laplacian and analogously for step (d) in transposed form) has the following Kronecker product form
\begin{equation}\label{eq:tensor_naive}
  \begin{bmatrix}
    D_{1} \otimes S_{2} \otimes \ldots \otimes S_{d}\\
    S_{1} \otimes D_{2} \otimes \ldots \otimes S_{d}\\
    \vdots\\
    S_{1} \otimes S_{2} \otimes \ldots \otimes D_{d}
  \end{bmatrix}
  \ve u^{(e)},
\end{equation}
and can therefore be implemented as a series of small matrix-matrix multiplications of dimension $k\times k$ for $S_i$ and $D_i$ and
$k\times k^{d-1}$ for the coefficients. These operations can
be interpreted as applying the matrices $S_{i}$ and $D_{i}$ along each
coordinate direction for $k^{d-1}$ lines. For evaluation of
Eq.~\eqref{eq:tensor_naive} in the collocation points, i.e., the case where
quadrature points for nodal polynomials coincide with the node positions, the
interpolation matrix is the $k\times k$ identity matrix, $S_{i}=I_i$. Omitting the calculations for interpolation is a classical optimization in spectral element codes
\cite{Deville02,Kopriva09}.

The collocation approach can be used to reduce
cost of \eqref{eq:tensor_naive} also for other bases: if we define a 1D
gradient matrix $\left(D_{i}^{\mathrm{co}}\right)_{q,j}$ as the gradient of Lagrange polynomials
$\varphi_{j}^\text{1D,co}(\xi_q)$ with nodes in the quadrature points, i.e., $D_i = D_i^{\mathrm{co}} S_i$, expression \eqref{eq:tensor_naive} can be rewritten as
\begin{equation}\label{eq:tensor_to_collocation}
  \underbrace{
    \begin{bmatrix}
      D_{1}^{\mathrm{co}} \otimes I_{2} \otimes \ldots \otimes I_{d}\\
      I_{1} \otimes D_{2}^{\mathrm{co}} \otimes \ldots \otimes I_{d}\\
      \vdots\\
      I_{1} \otimes I_{2} \otimes \ldots \otimes D_{d}^{\mathrm{co}}
    \end{bmatrix}
  }_{\text{collocation derivative}}
  \underbrace{
    \begin{bmatrix}
      S_{1} \otimes S_{2} \otimes \ldots \otimes S_{d}
    \end{bmatrix}
  }_{\text{basis change}}
  \ve u^{(e)},
\end{equation}
where the first multiplication
$\begin{bmatrix} S_{1} \otimes S_{2} \otimes \ldots \otimes S_{d}
\end{bmatrix} \ve u^{(e)}$ transforms the values $\ve u^{(e)}$ to the
nodal basis defined in quadrature points. This transformation reduces the number of tensor product calls for the gradient from $d^2$ in \eqref{eq:tensor_naive} to $2d$ on general bases or $4dk^{d+1}$ arithmetic operations per cell. For comparison, a naive implementation that does not utilize the tensor product structure would involve $2dk^{2d}$ arithmetic operations. Integration is performed by multiplication
with the transpose of the matrix in \eqref{eq:tensor_to_collocation}.

The interpolation matrix for face integrals is of similar form. For the case
of the DG Laplacian that uses both function values and derivatives of $u_h$, face integration first
interpolates values and gradients in normal direction to the face and then
applies $(d-1)$-dimensional tensor product operations. As an example, let us consider a face in 3D
with normal in $\xi_2$ direction. The interpolation matrix consists of four blocks---the first block for the values in the quadrature points, the two subsequent blocks for the
derivatives in the local coordinate direction of the face ($\xi_1$ and $\xi_3$
in this case), and the last block for the derivative in face-normal direction---and has the following structure
\begin{equation}\label{eq:face_integral}
  \underbrace{
    \begin{bmatrix}
      \begin{bmatrix}
    \begin{bmatrix}
      I_1 \otimes I_2\\
      D_{1}^{\mathrm{co}}\otimes I_{2}\\
      I_{1} \otimes  D_{2}^{\mathrm{co}}
    \end{bmatrix}
        \begin{bmatrix}
      S_{1} \otimes  S_{2}
    \end{bmatrix}\end{bmatrix}
    &  0 \\
    0 & \begin{bmatrix}S_{1}\otimes S_{2} \end{bmatrix}
  \end{bmatrix}
  }_{\text{interpolation within face}}
  \underbrace{
    \begin{bmatrix}
      I_{1} \otimes S_\text{f}  \otimes I_3 \\
      I_{1} \otimes D_\text{f} \otimes I_3
    \end{bmatrix}
  }_{\text{face-normal interpolation}}
  \ve u^{(e)},
\end{equation}
where the $1\times k$ matrices $S_\text{f}$ and $D_\text{f}$
evaluate the shape functions and their first derivative on the
respective boundary of the reference cell.
For derivatives in $\xi_1$ or $\xi_3$ directions the
interpolation matrices are moved to the respective slots in the face-normal
interpolation.

\begin{Remark}
  Note that we do not consider the case where the final cell matrix (steps
  (b)--(d) in Algorithm \ref{alg:prototype_dof_op}) can be represented as a
  sum of Kronecker products, such as the Laplacian on a Cartesian geometry
  that has the form $L = L_1\otimes M_2 + M_1\otimes L_2$ in 2D.  This
  involves fewer tensor product kernels than the numerical integration and can
  sometimes be further reduced in complexity \cite{huismann17}. These separable matrices
  only appear for the case of constant coefficients and axis-aligned meshes
  and are not the primary interest of this work, even though the presented sum
  factorization techniques can also be applied to that setting.
\end{Remark}

\subsection{Overview of algorithm design}

Algorithm \ref{alg:prototype_dof_op} has previously often been implemented by
specializations for the particular equations at hand or selected
parallelization schemes in the high performance community, which have been
discussed in an enormous body of literature on DG methods. This work attempts
a systematic approach to identify implementations that combine generic
software interfaces in C++ not tied to a particular equation with high performance execution.

The vector access steps (a) and (e) in
Algorithm \ref{alg:prototype_dof_op} only shuffle around memory and are thus memory bandwidth bound when considered in isolation. Conversely, the interpolation and integration steps (b) and (d) in Algorithm
\ref{alg:prototype_dof_op} repeatedly go through a limited set of data in different orders
and allow for caching: the data extracted from the input vector is combined
with the coefficients of the polynomial evaluation to some intermediate arrays holding the data on
all quadrature points, see Sec.~\ref{sec:kernel_local}. The operations performed in quadrature points, step (c)
in the algorithm, have most variability and depend on the differential
operator and the user code as discussed in Sec.~\ref{sec:geometry}. In order to assess the effects of code
optimizations separately, the following three sections discuss one of the
aspects with idealized implementations of the other steps of
Algorithm \ref{alg:prototype_dof_op}. Note that the presented optimization techniques do not interact with each other in the chosen order, so they can be derived one after another.

Given this initial characterization, it is clear that the memory-intensive operations in the
vector access and the geometry stages should be mixed with the compute-dominated interpolation and
integration steps such that the data that flows between
the different phases in the algorithm remains in cache memory as much as
possible and prefetchers can pre-load vector entries or coefficients during compute phases. Thus, implementations must perform all operations of a cell or
a few cells together with a single outer loop. For instance, a copy
of the vector data into formats more amenable to vectorization or basis change operations should be
done local to a few cells unless all data fits into the
last-level caches.
Along the same lines, the overall operator evaluation often performs best when interleaving
the cell loop (\ref{alg:prototype_cell}) with the inner face loop
(\ref{alg:prototype_face}) and the boundary face loop (\ref{alg:prototype_bound}) in Algorithm \ref{alg:prototype_dof_op} because they can re-use the
vector entries in caches.

\section{Performance-driven design of sum factorization kernels}\label{sec:kernel_local}

In this section, we start by an analysis of the sum factorization kernels for
cell integrals which have a complexity of $\mathcal O(k)$ per degree of
freedom.  To eliminate memory effects, we apply the kernels to data
corresponding to a single cell (or a batch of cells when vectorizing)
throughout this section. This data is repeatedly accessed in an outer loop
whose length is set to reach a benchmark run time of a few seconds. To give a
more realistic picture of application performance, we run the full cell
operation stage (b)--(d) in Algorithm~\ref{alg:prototype_dof_op} including the
loop over quadrature points that applies the geometry, without artificially
looking at the sum factorization step only. In particular, our numbers include
the cost of selecting between different code paths for curved or affine
geometries according to \cite{Kronbichler12}.

Our performance experiments are based on the three Intel-based
HPC systems presented in Table~\ref{tab:systems}, including two server
processors and a throughput-oriented Xeon Phi. We have verified that benchmarks
were run at the maximum AVX frequency given in the table, and report all
timings as the average of runs over at least a few seconds after a warm-up
period of a few minutes to represent as fair performance numbers as
possible. As a compiler we use the GNU compiler \texttt{g++} of version 6.3.0
with flags \texttt{-O3 -funroll-loops -march=native} that in our case generates
executables of slightly better performance than \texttt{clang} 3.9.0 and the
Intel compiler version 16 and 17.

The
Haswell system with $2\times 8$ cores is a system with relatively high memory
throughput as compared to arithmetics. The Broadwell system with $2\times 14$
cores at $2.6$ GHz has almost twice the core count and theoretically offers
twice the arithmetic performance of the Haswell system, but with the same
number of 8 DDR4 memory channels and only slightly higher STREAM memory
throughput due to a higher memory clock rate. The Knights Landing
Processor has 64 cores that run with wider vector units but at a low frequency of 1.3 GHz and with reduced
features, the most important of which are the only 2-wide decode throughput
and the absence of a level-3 cache.

\begin{table}
  \caption{Specification of hardware systems used for evaluation. Memory
    bandwidth according to the STREAM triad benchmark and GFLOP/s based on
    theoretical maximum at AVX turbo frequencies.}
  \label{tab:systems}
  \smallskip
  {
    \small\strut\hfill
    \begin{tabular}{lccc}
      \hline
      & Haswell  & Broadwell & Knights Landing\\
      & Xeon E5-2630 v3 & Xeon E5-2690 v4 & Xeon Phi 7120\\
      \hline
      cores       & $2\times 8$ & $2\times 14$ & $64$\\
      frequency base & 2.4 GHz & 2.6 GHz & 1.3 GHz\\
      max AVX turbo frequency & 2.6 GHz & 2.9 GHz & 1.1 GHz\\
      SIMD width    & 256 bit & 256 bit & 512 bit \\
      arithmetic peak @ AVX turbo & 666 GFLOP/s & 1299 GFLOP/s & 2253 GFLOP/s\\
      last level cache    & 2.5 MB/core L3  & 2.5 MB/core L3 & 512 kB/core L2\\
      memory bandwidth & 95 GB/s & 112 GB/s &  450 GB/s (MCDRAM)\\
      \hline
    \end{tabular}\hfill\strut
  }
\end{table}

\subsection{Tensor product algorithms}\label{sec:sum_factorization}

The summations in the interpolation kernel according to
Eq.~\eqref{eq:tensor_to_collocation} and its transpose for integration as used in Algorithm
\ref{alg:prototype_dof_op} involve a series of small matrix-matrix
multiplications and are the dominating factor for reaching high
performance.
Due to the small dimensions of the matrices of $k\times k$ for the coefficient array and $k\times k^2$ for the 3D solution values, generic BLAS multiplication kernels \texttt{dgemm} that are specialized for the LINPACK context of medium and large sizes are not suitable due to large function call and dispatch overheads. As an alternative, optimized small matrix multiplication kernels have
been suggested by the batched BLAS initiative \cite{Dongarra16} and by the
\texttt{libxsmm} project \cite{libxsmm17}. However, these interfaces potentially lose some of the context of the repeated matrix multiplications along different directions, i.e., with different strides, leading to additional load/store operations between registers and the L1 cache that are often the limit for throughput.
Also, separating the tensor product kernels from
the quadrature loop and vector access as assumed by batched BLAS
\cite{Dongarra16} sacrifices data locality and thus
performance in the low and medium polynomial degree case.
Instead, this setting is best handled by an
optimizing compiler in our experience.
Obviously,
enough information about the loops and data arrays must be given to the
compiler to ensure generation of good machine code. As we will show below,
very good numbers can indeed be obtained with compiled C++ code when letting the
compiler choose the loop unrolling and register allocation, but forcing the use of full SIMD vectors by intrinsics.

For the steps (b) and (d) in Algorithm \ref{alg:prototype_dof_op}, we consider three compute kernels, \texttt{basis\_change}, \texttt{collocation\_derivative}, and \texttt{face\_normal\_interpolation}. Before discussing their optimization, let us briefly introduce the properties of the kernels and their signature.

\begin{center}
  \small
  \begin{minipage}{0.92\textwidth}
\begin{verbatim}
template <int d, bool interpolate, int k, int l, bool add, typename Number>
void basis_change(const Number* shape_values, const Number* in, Number* out)
\end{verbatim}
  \end{minipage}
\end{center}

\noindent The \texttt{basis\_change} kernel implements the change between different polynomial bases in $d$
space dimensions. It involves $d$ tensor product sweeps passing through each
dimension according to the matrix
$[S_1\otimes S_2\otimes\ldots\otimes S_d]$ in
Eq.~\eqref{eq:tensor_to_collocation}, with the array \texttt{shape\_values}
storing the entries of the $S$ matrices. The boolean template argument
\texttt{interpolate} selects either the interpolation or the integration path. The algorithm is generic, though, and the
switch between interpolation and integration is merely between interpreting the
shape value array as either row-major or column-major. The 1D size of the two bases is given by $k$
on the degree of freedom dimension and $l$ on the quadrature dimension,
respectively. Well-defined interpolation must satisfy $k\leq l$, i.e.,
interpolate to a basis of the same degree or a higher degree. The boolean
template argument \texttt{add} defines whether the result should be added into
the output array or overwrite previous content. This is used when adding the
local results of an integral into previous content in a global vector, for
instance. The result can be stored in place if the data along the 1D input
stripe is held in registers while passing through the rows of the 1D
interpolation matrix and appropriate order through the input array.
The number type \texttt{Number} is either a simple scalar like
\texttt{double} or a more complicated SIMD type that we call
\texttt{VectorizedArray<double>} or \texttt{VectorizedArray<float>} with
overloaded arithmetic operations.

For a mass matrix evaluation or an inverse mass matrix evaluation according to
\cite{Kronbichler16}, the basis change kernel is first applied in the
\texttt{interpolate=true} version, the determinant of the Jacobian and the
quadrature weights are applied in each quadrature point (multiplication by a diagonal
matrix), and integration is performed by calling the kernel with
\texttt{interpolate=false}, using the same \texttt{shape\_values} array as for interpolation.

\begin{center}
  \small
  \begin{minipage}{0.92\textwidth}
\begin{verbatim}
template <int d, bool interpolate, int k, bool add, typename Number>
void collocation_derivative(const Number* shape_derivatives, const Number* in,
                            Number* out)
\end{verbatim}
  \end{minipage}
\end{center}

\noindent The \texttt{collocation\_derivative} kernel implements the unit-cell derivative in collocation space
according to the left matrix in Eq.~\eqref{eq:tensor_to_collocation} using $d$
tensor product kernel invocations on $k^d$ points each. If
\texttt{interpolate==true}, $d$ components of the gradient are created from a
single input field, whereas $d$ consecutive fields are summed together if \texttt{interpolate==false}.

For evaluating the unit-cell gradient according to
Eq.~\eqref{eq:tensor_to_collocation} on an arbitrary basis, the two kernels
\texttt{basis\_change} and \texttt{collocation\_derivative} are applied
sequentially for the same cell batch, involving $2d$ tensor product kernels. Of course, our algorithm
design can bypass \texttt{basis\_change} for collocation of nodal points and
quadrature points. Algorithms similar to \texttt{basis\_change} and
\texttt{collocation\_derivative} were already proposed in \cite{Kopriva09} but
without discussion of their properties of modern hardware.

\begin{center}
  \small
  \begin{minipage}{0.92\textwidth}
\begin{verbatim}
template <int d, int direction, bool interpolate, int k, int highest_derivative,
          typename Number>
void face_normal_interpolation(const Number* shape_matrix, const Number* in,
                               Number* out)
\end{verbatim}
  \end{minipage}
\end{center}

\noindent This kernel implements the interpolation of values and gradients in
face-normal direction (right part of Eq.~\eqref{eq:face_integral}) when passing either $S_\text{f}$ or $D_\text{f}$ as \texttt{shape\_matrix}. It contains short-cuts for special polynomial bases where most of the entries in $S_\text{f}$ and $D_\text{f}$ are zero: If only a single entry in $S_\text{f}\in \mathbb{R}^{1\times k}$ is non-zero, only $k^{d-1}$ coefficient values contribute to the interpolation on the face (\texttt{nodal\_on\_faces}). Likewise, a Hermite-type basis (\texttt{hermite\_type\_basis}) where at most two entries in both $S_\text{f}$ and $D_\text{f}$ are nonzero, respectively, allows to compute values and derivatives from only $2k^{d-1}$ coefficients. This functionality can be integrated with the vector access functions
according to Sec.~\ref{sec:vector_access} below to skip reading the remaining vector entries. The interpolation and
integration steps (b) and (d) within the face are performed by combining the
\texttt{face\_normal\_interpolation} kernel with the $(d-1)$-dimensional version
of the \texttt{basis\_change} and \texttt{collocation\_derivative} functions.
All components of face integrals involve only
$\mathcal O(1)$ arithmetic operations per cell degree of freedom. They reduce in complexity against
cell integrals at higher convergence orders $k$ and we postpone the
analysis of face integrals to subsection~\ref{sec:cells_and_faces}.

\begin{table}
  \caption{Number of calls to tensor product kernels in $d$ dimensions for
    the advection operator with a Lagrange basis on Gauss--Lobatto points and the
    Laplacian on a Hermite-type basis using Gaussian quadrature on $k^d$ points. The basis change and derivative columns specify how the total
    number of sum factorization kernels is derived.}
  \label{tab:tensor_costs}
  \smallskip
  {
    \small\strut\hfill
    \begin{tabular}{lllll}
      \hline
      & total no.~of kernels & basis change & derivative & face normal \\
      \hline
      advection GL, cell & $3d$ on $k^d$ pts & $2d$ & $d$ & ---\\
      advection GL, inner face & $4d$ on $k^{d-1}$ pts & $4(d-1)$ & --- & 4 on $k^{d-1}$ pts\\
      advection GL, boundary face & $2d$ on $k^{d-1}$ pts & $2(d-1)$ & --- & 2 on $k^{d-1}$ pts\\
      \hline
      Laplacian Hermite, cell & $4d$ on $k^d$ pts & $2d$ & $2d$ & --- \\
      Laplacian Hermite, inner face & $12(d-1)$ on $k^{d-1}$ pts & $8(d-1)$ & $4(d-1)$ & 8 on $2k^{d-1}$ pts\\
      Laplacian Hermite, bdry face & $6(d-1)$ on $k^{d-1}$ pts & $4(d-1)$ & $2(d-1)$ & 4 on $2k^{d-1}$ pts\\
      \hline
    \end{tabular}
    \hfill\strut
  }
\end{table}

Table~\ref{tab:tensor_costs} specifies the number of tensor product
invocations when calling Algorithm~\ref{alg:prototype_dof_op} for the
advection operator on the left hand side of \eqref{eq:advection} and the respective
algorithm for the left hand side of the Laplacian \eqref{eq:poisson_sip}. A nodal basis on Gauss--Lobatto points and a Hermite-like basis, respectively, are chosen such that the operator evaluation time is minimal for polynomial degrees up to 10. We explicitly note that the operator evaluation with the collocation basis executes more slowly on an actual implementation despite fewer tensor product calls. This is because \texttt{face\_normal\_interpolation} must access all $k^d$ points of a
cell rather than only the $2k^{d-1}$ with non-zero value and first derivative on the face. A comparative analysis will be performed in a subsequent contribution. The table reports
the total number of tensor product kernels for both the interpolation step (b) and the
integration step (d).
The
higher order of the derivatives in the Laplacian imply more work, in
particular for the face integrals where both the values and the gradients must
be computed from Eq.~\eqref{eq:face_integral}.
In
Table~\ref{tab:tensor_costs}, we count the interpolation of the values and
normal derivatives as two invocations to a face normal interpolation to
quantify the increased cost, even though they are implemented by
a single pass through the data.

\subsection{Vectorization strategy}\label{sec:vectorize}

Modern high-performance CPU architectures increasingly rely on
single-instruction/multi\-ple-data (SIMD) primitives as a means to improve
performance per watt. In short-vector SIMD, a single arithmetic or
load/store operation is issued to process a number of $n_\text{lanes}$ data
elements with the same instruction independently. Cross-lane permutations
require separate instructions that may incur a performance penalty, depending on
the superscalar execution capabilities of the microarchitecture. Furthermore,
loads and stores are faster if accessing a contiguous range of memory
(packed operation) as compared to indirect addressing with \texttt{gather}
or \texttt{scatter} instructions with multiple address generation steps.

A first, most obvious, option is to vectorize within a cell clustering 1D slices of the
degrees of freedom. For sum factorization, this is intrigued
because the 1D kernels go through data in different orders for each
dimension. As visualized in Fig.~\ref{fig:vectorization_within_cells}, three
lane-crossing transpose operations of length $k^d$ are needed in the spirit of array-of-struct into struct-of-array conversions.
In case the number of 1D degrees of freedom is not a
multiple of the SIMD width, one needs to fill up the last SIMD array along that
direction with dummy values, which leads to distinct drops in the performance.
Alternatively, smaller SIMD widths could be used for the remainder parts.

\begin{figure}
  \centering
  \begin{tikzpicture}[scale=0.6]
    \draw[->] (-6.5,-1) -- (-5.5,-1) node [right] {\small $x$};
    \draw[->] (-6.5,-1) -- (-6.5,0) node [above] {\small $z$};
    \node at (-4,2.8) {\scriptsize Values in nodes};
    \foreach \y in {0,0.55,1.45,2}
      \draw[color=red!50!white,fill=red!20!white,rounded corners=2] (-5.2,\y-0.2) rectangle (-2.8,\y+0.2);
    \draw[thick] (-5,0) rectangle (-3,2);
    \foreach \x in {-5,-4.45,-3.55,-3}
    \foreach \y in {0,0.55,1.45,2}
    \draw[black,fill=black] (\x,\y) circle (0.1);
    \draw[blue,ultra thick,<->] (-2.3,0.) -- node[right]{$z$} (-2.3,2);

    \node at (1,2.8) {\scriptsize Values in $q$-points};
    \foreach \x in {0.14,0.66,1.34,1.86}
      \draw[color=red!50!white,fill=red!20!white,rounded corners=2] (\x-0.2,-0.2) rectangle (\x+0.2,2.2);
    \draw[thick] (0,0) rectangle (2,2);
    \foreach \x in {0.14,0.66,1.34,1.86}
    \foreach \y in {0.14,0.66,1.34,1.86}
    \draw[black,fill=black] (\x,\y) circle (0.1);
    \draw[blue,ultra thick,<->] (0.,-1.2) -- node[above]{$x$} (2,-1.2);

    \draw[blue,ultra thick,<->] (-0.8,-1.2) circle (0.02);
    \draw[blue,ultra thick,<->] (-0.8,-1.2) circle (0.25) node[above=0.1]{$y$};

    \node at (5,2.8) {\scriptsize $\frac{\partial u_h}{\partial x}$};
    \foreach \x in {4.14,4.66,5.34,5.86}
      \draw[color=red!50!white,fill=red!20!white,rounded corners=2] (\x-0.2,-0.2) rectangle (\x+0.2,2.2);
    \draw[thick] (4,0) rectangle (6,2);
    \foreach \x in {4.14,4.66,5.34,5.86}
    \foreach \y in {0.14,0.66,1.34,1.86}
    \draw[black,fill=black] (\x,\y) circle (0.1);
    \draw[blue,ultra thick,<->] (4.,-1.2) -- node[above]{$x$} (6,-1.2);

    \node at (9,2.8) {\scriptsize $\frac{\partial u_h}{\partial y}$};
    \foreach \x in {8.14,8.66,9.34,9.86}
      \draw[color=red!50!white,fill=red!20!white,rounded corners=2] (\x-0.2,-0.2) rectangle (\x+0.2,2.2);
    \draw[thick] (8,0) rectangle (10,2);
    \foreach \x in {8.14,8.66,9.34,9.86}
    \foreach \y in {0.14,0.66,1.34,1.86}
    \draw[black,fill=black] (\x,\y) circle (0.1);

    \draw[blue,ultra thick,<->] (9,-1.2) circle (0.02);
    \draw[blue,ultra thick,<->] (9,-1.2) circle (0.25) node[above=0.02]{$y$};

    \node at (13,2.8) {\scriptsize $\frac{\partial u_h}{\partial y}$};
    \foreach \y in {0.14,0.66,1.34,1.86}
      \draw[color=red!50!white,fill=red!20!white,rounded corners=2] (11.8,\y-0.2) rectangle (14.2,\y+0.2);
    \draw[thick] (12,0) rectangle (14,2);
    \foreach \x in {12.14,12.66,13.34,13.86}
    \foreach \y in {0.14,0.66,1.34,1.86}
    \draw[black,fill=black] (\x,\y) circle (0.1);
    \draw[blue,ultra thick,<->] (14.7,0) -- node[right]{$z$} (14.7,2);
  \end{tikzpicture}
  \caption{Visualization of a possible vectorization scheme within a single
    element shown in $x-z$ plane with $y$-direction orthogonal to the sketch. Red
    shaded symbols denote the entries within a single SIMD array, involving
    transpose (cross-lane) operations to switch from one order to the
    other. Blue arrows show the action of 1D interpolations, one arrow per
    tensor product invocation.}
  \label{fig:vectorization_within_cells}
\end{figure}
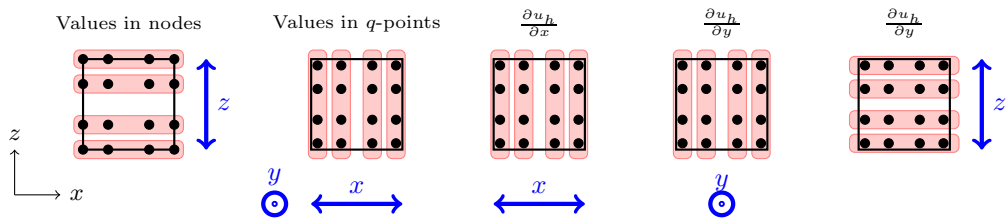

\begin{figure}
  \centering
  \begin{tikzpicture}[scale=1]
    \foreach \y in {0,2}
    \foreach \x in {0,2,4}
    \draw[green!20!white,fill=green!20!white] (\x+0.05,\y+0.05) rectangle (\x+1.95,\y+1.95);
    \foreach \y in {0,2,4}
    {
    \draw(0,\y) -- (6.2,\y);
    \node at (6.5,\y) {\ldots};
    }
    \foreach \x in {0,2,4,6}
    \draw(\x,0) -- (\x,4);
    \foreach \y in {0,1,2,3}
    \foreach \x in {0,1,2,3}
    {
      \pgfmathsetmacro\xx{0.25+0.5*\x}
      \pgfmathsetmacro\yy{0.25+0.5*\y}
      \pgfmathsetmacro\result{4*(\x+4*\y)}
      \node at(\xx,\yy){\scriptsize \pgfmathprintnumber{\result}};
    }
    \foreach \y in {0,1,2,3}
    \foreach \x in {0,1,2,3}
    {
      \pgfmathsetmacro\xx{2+0.25+0.5*\x}
      \pgfmathsetmacro\yy{0.25+0.5*\y}
      \pgfmathsetmacro\result{4*(\x+4*\y)+1}
      \node at(\xx,\yy){\scriptsize \pgfmathprintnumber{\result}};
    }
    \foreach \y in {0,1,2,3}
    \foreach \x in {0,1,2,3}
    {
      \pgfmathsetmacro\xx{0.25+0.5*\x}
      \pgfmathsetmacro\yy{2+0.25+0.5*\y}
      \pgfmathsetmacro\result{4*(\x+4*\y)+2}
      \node at(\xx,\yy){\scriptsize \pgfmathprintnumber{\result}};
    }
    \foreach \y in {0,1,2,3}
    \foreach \x in {0,1,2,3}
    {
      \pgfmathsetmacro\xx{2+0.25+0.5*\x}
      \pgfmathsetmacro\yy{2+0.25+0.5*\y}
      \pgfmathsetmacro\result{4*(\x+4*\y)+3}
      \node at(\xx,\yy){\scriptsize \pgfmathprintnumber{\result}};
    }
    \foreach \y in {0,1,2,3}
    \foreach \x in {0,1,2,3}
    {
      \pgfmathsetmacro\xx{4.25+0.5*\x}
      \pgfmathsetmacro\yy{0.25+0.5*\y}
      \pgfmathsetmacro\result{4*(\x+4*\y)+64}
      \node at(\xx,\yy){\scriptsize \pgfmathprintnumber{\result}};
    }
    \foreach \y in {0,1,2,3}
    \foreach \x in {0,1,2,3}
    {
      \pgfmathsetmacro\xx{4.25+0.5*\x}
      \pgfmathsetmacro\yy{2.25+0.5*\y}
      \pgfmathsetmacro\result{4*(\x+4*\y)+66}
      \node at(\xx,\yy){\scriptsize \pgfmathprintnumber{\result}};
    }
    \node at (1,-0.25)[green!50!black] {$e_1$};
    \node at (3,-0.25)[green!50!black] {$e_2$};
    \node at (5,-0.25)[green!50!black] {$e_5$};
    \node at (1,4.25)[green!50!black] {$e_3$};
    \node at (3,4.25)[green!50!black] {$e_4$};
    \node at (5,4.25)[green!50!black] {$e_7$};
  \end{tikzpicture}
  \caption{Visualization of layout of degrees of freedom for vectorization
    over elements, using a typical array-of-structure-of-array data
    structure.}
  \label{fig:vectorization_over_cells}
\end{figure}

The second option, which is preferable in our experience, is to vectorize over several
cells. Fig.~\ref{fig:vectorization_over_cells} shows a numbering of degrees of
freedom on a $\mathcal Q_3$ basis in 2D with 4-wide vectorization which
allows for direct packed access. In this layout, the lower left node values of
four cells are placed adjacent in memory. The next storage location is the
second node for all four cells, and so on.
In this format, no cross-lane data exchange
is needed for cell integrals and the sum factorization kernels can be directly
called on the data stored in the global input vector without a separate gather step. When vectorizing over several cells,
all arithmetic operations can be performed straight-forwardly, including the various tensor
product kernels and the operations in quadrature points, by using overloaded
SIMD data types according to
\cite{Kronbichler12,Kronbichler17}. Furthermore, this scheme can straight-forwardly
select the most beneficial width of vectorization for a given hardware. Partially filled lanes occur
at most on a single element batch per operator evaluation and MPI rank for
meshes whose number of cells is not divisible by the
vectorization width. We also apply this approach to face integrals,
i.e., we process the integrals of several faces at once, rather than
SIMD-parallelizing within a face or over the two cells adjacent to a face.

In order to maintain data locality, we finish all
operations of steps (a)--(e) in Algorithm~\ref{alg:prototype_dof_op} on a batch of cells, e.g. the cells indexed by
$\{e_1, e_2, e_3, e_4\}$ in Fig.~\ref{fig:vectorization_over_cells}, before
proceeding with the next batch of cells, e.g. cells
$\{e_5,e_6,e_7,e_8\}$.
Besides requiring a change in the loop over the mesh, two possible disadvantages of this scheme are
\begin{itemize}
\item a somewhat
larger spread in the indices of gather/scatter steps of face integrals,
\item cases where the number of elements per MPI task is less than the width
  of vectorization do not see speedups, which is usually only limiting over
  the communication cost for more than 500 degrees of freedom per cell
  \cite{Kronbichler16b}, and
\item a larger active size of the temporary arrays in sum factorization, which
  might exceed the capacity of caches and thus slow down execution. In the
  following, we show that this issue is uncritical in the common context
  of $p< 15$, i.e., $k< 16$.
\end{itemize}

\begin{figure}
  \centering
  \begin{tikzpicture}
    \begin{semilogyaxis}[
      width=0.51\textwidth,
      height=0.42\textwidth,
      xlabel={Polynomial degree $p$},
      ylabel={Degrees of freedom per second},
      tick label style={font=\scriptsize},
      label style={font=\scriptsize},
      title style={at={(1,0.952)},anchor=north east,draw=black,fill=white,font=\scriptsize},
      title={2D Laplacian},
      xmin=0,xmax=25,
      ymin=2e8,ymax=1e10,
      grid,
      cycle list name=colorGPL,
      mark size=1.8,
      semithick
      ]
      \addplot[gnuplot@darkblue,mark=square] table[x={deg}, y={VectUrTwo}] {\cellHaswellVecCompareThree};
      \addplot[red!80!white,mark=o] table[x={deg}, y={VectNrTwo}] {\cellHaswellVecCompareThree};
      \addplot[gnuplot@orange,mark=diamond*] table[x={deg}, y={NoVectTwo}] {\cellHaswellVecCompareThree};
    \end{semilogyaxis}
  \end{tikzpicture}
  \
  \begin{tikzpicture}
    \begin{semilogyaxis}[
      width=0.51\textwidth,
      height=0.42\textwidth,
      xlabel={Polynomial degree $p$},
      tick label style={font=\scriptsize},
      label style={font=\scriptsize},
      xmin=0,xmax=25,
      ymin=2e8,ymax=1e10,
      title style={at={(1,0.952)},anchor=north east,draw=black,fill=white,font=\scriptsize},
      title={3D Laplacian},
      legend to name=legendVectorization,
      legend columns=2,
      legend style={font=\scriptsize},
      grid,
      cycle list name=colorGPL,
      mark size=1.8,
      semithick
      ]
      \addplot[gnuplot@darkblue,mark=square] table[x={deg}, y={VectUrThree}] {\cellHaswellVecCompareThree};
      \addlegendentry{vectorized over cells tiled};
      \addplot[red!80!white,mark=o] table[x={deg}, y={VectNrThree}] {\cellHaswellVecCompareThree};
      \addlegendentry{vectorized over cells plain};
      \addplot[black,every mark/.append style={fill=green!80!black},mark=otimes*] table[x={deg}, y={VectEleThree}] {\cellHaswellVecCompareThree};
      \addlegendentry{vectorized within cells};
      \addplot[gnuplot@orange,mark=diamond*] table[x={deg}, y={NoVectThree}] {\cellHaswellVecCompareThree};
      \addlegendentry{auto-vectorization only};
    \end{semilogyaxis}
  \end{tikzpicture}
  \\
  \ref{legendVectorization}
  \caption{Comparison of throughput of local cell kernels on fully populated
    $2\times 8$ cores Intel Haswell (Xeon E5-2630 v3 @ 2.4 GHz) for cell
    integrals with respect to vectorization for Laplacian
    in 2D and 3D.}
\label{fig:cell_ops_vectorization}
\end{figure}
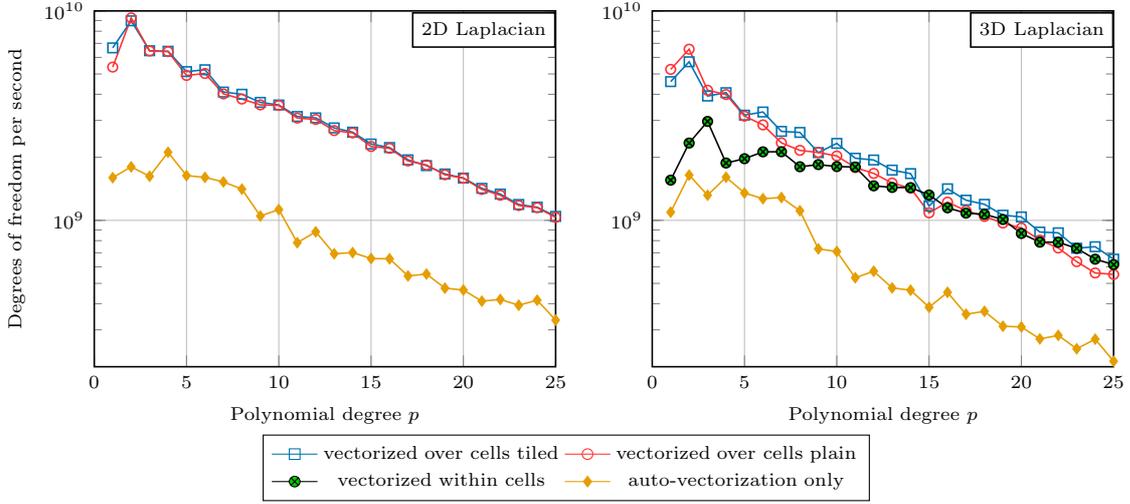

The performance of the two
vectorization variants, implemented through wrappers around
intrinsics, is shown in Fig.~\ref{fig:cell_ops_vectorization}. The figure also contains data points with automatic vectorization as exploited by the GNU
compiler with \texttt{\_\_restrict} annotations to arrays
to help the compiler in the aliasing analysis. Vectorization
over several cells clearly outperforms vectorization within a single cell for low and moderate polynomial degrees $1\leq p \leq 10$. While
the former shows an approximately linear degradation of performance due to the
complexity $\mathcal O(p+1)$ in the cell integrals, the latter increases
throughput as the lanes get more populated, with sudden drops in performance
once an additional lane must be used at $p=4, 8, 12, 16, 20, 24$. The
automatic vectorization is not competitive, with a performance disadvantage of
a factor 2.1 to 3.7 (average in 3D: 3.1) with 4-wide vectorization. This is
because only 5\% to at most 25\% of arithmetic instructions are done in packed
form.

In order to improve data locality of the sum factorization, Algorithm
\ref{alg:kernel_tiling} proposes to merge loops over $x$ and $y$ within a single loop over the last direction $z$. This exploits the natural cache blocking of tensor products and is particularly useful for vectorization over several cells which have a larger local data set than the scalar variant.
The corresponding data point is marked as
``vectorized over cells tiled'' in Fig.~\ref{fig:cell_ops_vectorization}.

\begin{algorithm}
  \caption{Loop tiling for sum-factorized evaluation of the cell Laplacian in 3D}
  \label{alg:kernel_tiling}
  \begin{itemize}
  \setlength{\itemsep}{0pt}
  \setlength{\parsep}{0pt}
  \setlength{\parskip}{0pt}
  \item for $i_z=0,\ldots,k-1$
    \begin{itemize}
  \setlength{\itemsep}{0pt}
  \setlength{\parsep}{0pt}
  \setlength{\parskip}{0pt}
    \item Kernel for $S_1$ along $x$ for $k^2$ points in $xy$ plane with
      offset $(i_z-1)k^2$ and stride $1$
    \item Kernel for $S_2$ along $y$ for $k^2$ points in $xy$ plane with
      offset $(i_z-1)k^2$ and stride $k$
    \end{itemize}
  \item for $i=0,\ldots,k^2-1$
    \begin{itemize}
  \setlength{\itemsep}{0pt}
  \setlength{\parsep}{0pt}
  \setlength{\parskip}{0pt}
    \item Kernel for $S_3$ along $z$ for $k$ points with offset $i$ and stride $k^2$
    \item Kernel for $D_3^{\text{co}}$ along $z$ for $k$ points with offset
      $i$ and stride $k^2$
    \end{itemize}
  \item for $i_z=0,\ldots,k-1$
    \begin{itemize}
  \setlength{\itemsep}{0pt}
  \setlength{\parsep}{0pt}
  \setlength{\parskip}{0pt}
    \item Kernel for $D_2^{\text{co}}$ along $y$ for $k^2$ points with offset
      $i_zk^2$
    \item for $i_y=0,\ldots,k-1$
      \begin{itemize}
  \setlength{\itemsep}{0pt}
  \setlength{\parsep}{0pt}
  \setlength{\parskip}{0pt}
      \item Kernel for $D_1^{\text{co}}$ along $x$ for $k$ points with offset
        $i_zk^2+i_yk$
      \item Apply Laplacian on $k$ quadrature points along $x$ direction with
        offset $i_zk^2+i_yk$ according to step (c) of
        Algorithm \ref{alg:prototype_dof_op}.
      \item Kernel for $\left(D_1^{\text{co}}\right)^\mathrm{T}$ (integration)
        along $x$ for $k$ points with offset $i_zk^2+i_yk$
      \end{itemize}
    \item Kernel for $\left(D_2^{\text{co}}\right)^\mathrm{T}$ along $y$ for
      $k^2$ points with offset $i_zk^2$; sum into result from $x$ direction
    \end{itemize}
  \item for $i=0,\ldots,k^2-1$
    \begin{itemize}
  \setlength{\itemsep}{0pt}
  \setlength{\parsep}{0pt}
  \setlength{\parskip}{0pt}
    \item Kernel for $\left(D_3^{\text{co}}\right)^\mathrm{T}$ along $z$ for
      $k$ points with offset $i$; sum into results from $x,y$
      directions
    \item Kernel for $S_3^\mathrm{T}$ along $z$ for $k$ points with offset $i$
      and stride $k^2$
    \end{itemize}
  \item for $i_z=0,\ldots,k-1$
    \begin{itemize}
  \setlength{\itemsep}{0pt}
  \setlength{\parsep}{0pt}
  \setlength{\parskip}{0pt}
    \item Kernel for $S_2^{\mathrm{T}}$ along $y$ for
      $k^2$ points in $xy$ plane with
      offset $i_zk^2$ and stride $k$
    \item Kernel for $S_1^{\mathrm{T}}$ along $x$ for $k^2$ points in
      $xy$ plane with offset $(i_z-1)k^2$ and stride $1$
    \end{itemize}
  \end{itemize}
\end{algorithm}

\begin{figure}
  \centering
  \begin{tikzpicture}
    \begin{axis}[
      width=0.6\textwidth,
      height=0.45\textwidth,
      xlabel={Polynomial degree $p$},
      ylabel={Bytes / DoF},
      tick label style={font=\scriptsize},
      label style={font=\scriptsize},
      legend to name=legendCaches,
      legend columns=3,
      legend style={font=\scriptsize},
      xmin=0,xmax=25,
      ymin=0,ymax=400,
      grid,
      cycle list name=colorGPL,
      mark size=1.8,
      semithick
      ]
      \addplot[draw=gnuplot@darkblue,mark=square] table[x expr={\thisrowno{0}}, y expr={\thisrowno{1}+\thisrowno{2}}] {\cellCacheAccessNewUnrolled};
      \addlegendentry{vectorized over cells tiled L1 $\leftrightarrow$ L2 transfer};
      \addplot[draw=gnuplot@darkblue,densely dashed,mark=square,every mark/.append style={solid}] table[x expr={\thisrowno{0}}, y expr={\thisrowno{3}+\thisrowno{4}}] {\cellCacheAccessNewUnrolled};
      \addlegendentry{L2 $\leftrightarrow$ L3 transfer};
      \addplot[draw=gnuplot@darkblue,densely dotted,mark=square,every mark/.append style={solid}] table[x expr={\thisrowno{0}}, y expr={\thisrowno{5}+\thisrowno{6}}] {\cellCacheAccessNewUnrolled};
      \addlegendentry{L3 $\leftrightarrow$ RAM transfer};
      \addplot[draw=red!80!white,mark=o,every mark/.append style={solid}] table[x expr={\thisrowno{0}}, y expr={\thisrowno{1}+\thisrowno{2}}] {\cellCacheAccessNewNotUnrolled};
      \addlegendentry{vectorized over cells plain L1 $\leftrightarrow$ L2 transfer};
      \addplot[draw=red!80!white,densely dashed,mark=o,every mark/.append style={solid}] table[x expr={\thisrowno{0}}, y expr={\thisrowno{3}+\thisrowno{4}}] {\cellCacheAccessNewNotUnrolled};
      \addlegendentry{L2 $\leftrightarrow$ L3 transfer};
      \addplot[draw=red!80!white,densely dotted,mark=o,every mark/.append style={solid}] table[x expr={\thisrowno{0}}, y expr={\thisrowno{5}+\thisrowno{6}}] {\cellCacheAccessNewNotUnrolled};
      \addlegendentry{L3 $\leftrightarrow$ RAM transfer};
      \addplot[draw=black,mark=otimes*,every mark/.append style={solid,fill=green!80!black}] table[x expr={\thisrowno{0}}, y expr={\thisrowno{1}+\thisrowno{2}}] {\cellCacheAccessVectWithinCells};
      \addlegendentry{vectorized within cell L1 $\leftrightarrow$ L2 transfer};
      \addplot[draw=black,mark=otimes*,densely dashed,every mark/.append style={solid,fill=green!80!black}] table[x expr={\thisrowno{0}}, y expr={\thisrowno{3}+\thisrowno{4}}] {\cellCacheAccessVectWithinCells};
      \addlegendentry{L2 $\leftrightarrow$ L3 transfer};
      \addplot[draw=black,mark=otimes*,densely dotted,every mark/.append style={solid,fill=green!80!black}] table[x expr={\thisrowno{0}}, y expr={\thisrowno{5}+\thisrowno{6}}] {\cellCacheAccessVectWithinCells};
      \addlegendentry{L3 $\leftrightarrow$ RAM transfer};
    \end{axis}
  \end{tikzpicture}
  \\
  \ref{legendCaches}
  \caption{Memory access per degree of freedom on various levels of the memory
    hierarchy for the compute part of the 3D Laplacian run on all cores of an
    Intel Haswell Xeon E5-2630 v3.}
\label{fig:cell_memory_access}
\end{figure}
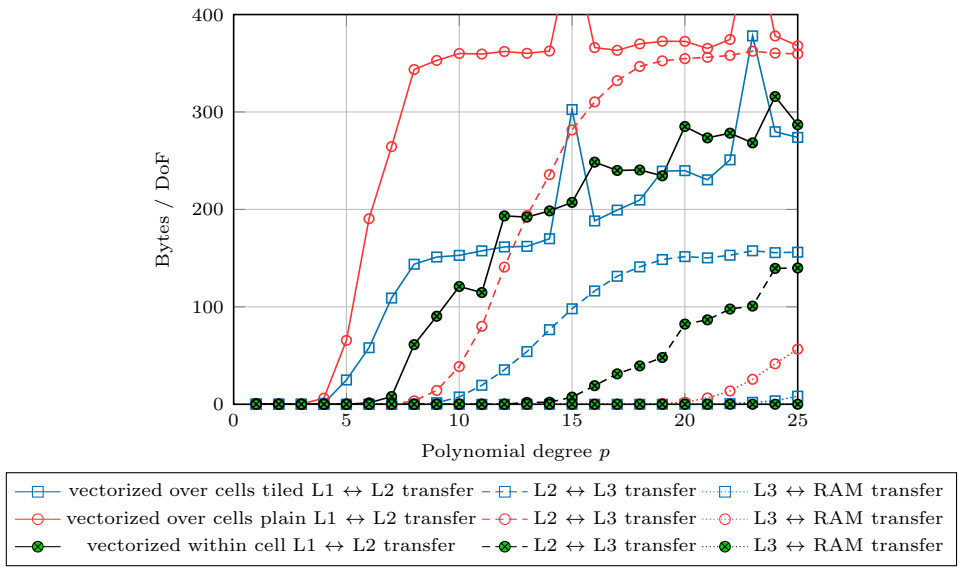

Further details on the kernel variants are provided by a cache access analysis that repeatedly runs through the kernels on a single batch of cells. Fig.~\ref{fig:cell_memory_access} shows
measurements of the actual data transfer (read + evict) between the various cache levels of
an Intel Haswell Xeon E5-2630 v3 with 4-wide vectorization extracted from
hardware performance counters with the \texttt{likwid}\footnote{https://github.com/rrze-hpc/likwid, retrieved on May 20,
  2017.} tool with \texttt{likwid-mpirun -n
  16 -g CACHES} to populate all 16 cores of the system. Raising the polynomial
degree increases the size of the temporary arrays holding intermediate results of sum factorization as expected. For vectorization over cells, degrees larger than 5
start spilling to L2 cache and degrees larger than 10 spill to L3
cache. The tiled algorithm cuts data access into a half to a third for larger polynomial degrees because data is aggregated along 2D planes. For
vectorization within cells, the active set is smaller by a factor of four
approximately, and spilling to L1 only starts
at degree 8 and spilling to L3 cache gets significant only for degrees larger
than 20.

However, the reduced transfer from caches when vectorizing within cells does
not materialize in better performance. This is because more transfer from
outer level caches comes along with more arithmetics due to the linear
increase in work according to the complexity $\mathcal O(k)$ per DoF, nicely
offsetting the reduced throughput and increased latency of outer level
caches. When compared to the access to L1 cache which is around 800--2000
Bytes/DoF (larger numbers for larger degrees), access of up to 400 Bytes/DoF
hitting the L2 cache is uncritical for vectorization over cells: the L2 cache
can sustain about half to one third of the throughput of the L1 cache
\cite{IntelOptimizationManual}. Likewise, throughput of the L3 cache is around
half that of L2 cache in recent Intel architectures, which is again a nice fit
with the access patterns in sum factorization according to
Fig.~\ref{fig:cell_memory_access} (square symbols).

Vectorization within cells only
becomes superior beyond $k=21$ in 3D where already the size of the local
vector data is
$21^3\, [\text{DoFs per cell}] \times 32\, [\text{Bytes/AVX SIMD}] = 296\,
\text{kiB}$ and thus exceeds the level 2 cache on Intel processors with AVX
vectorization. The only apparent outlier is degree $15$ with $k=16$ on
$16^3$ degrees of freedom per cell which is affected by cache associativity
limitations due to access to $64$ entries at a distance of exactly
$8\,\text{kiB} = 256 \times 32[\text{Bytes}]$ in between.

\begin{Requirement}
  In operator evaluation, quadrature point operations according to step~(c) in Algorithm~\ref{alg:prototype_dof_op} can take a substantial amount of time
  compared to sum factorization, in particular for face integrals. To obtain
  optimal performance, vectorization must also be applied to the operation in
  the quadrature points. Vectorization can be encapsulated in user code by
  vectorized data types and operator overloading.
\end{Requirement}

\subsection{Optimization of loop kernels}\label{sec:loop_kernels}

\begin{figure}
  \centering
  \begin{tikzpicture}
    \begin{semilogyaxis}[
      width=0.51\textwidth,
      height=0.42\textwidth,
      xlabel={Polynomial degree $p$},
      ylabel={Degrees of freedom per second},
      tick label style={font=\scriptsize},
      label style={font=\scriptsize},
      legend to name=legend2DHasw,
      legend columns=3,
      legend style={font=\scriptsize},
      title style={at={(1,0.952)},anchor=north east,draw=black,fill=white,font=\scriptsize},
      title={2D Laplacian},
      xmin=0,xmax=25,
      ymin=1e8,ymax=1e10,
      grid,
      cycle list name=colorGPL,
      mark size=1.8,
      semithick
      ]
      \addplot table[x={deg}, y={EvenOdd}] {\cellHaswellLoopsTwoD};
      \addlegendentry{templated, even-odd};
      \addplot table[x={deg}, y={UnrollFour}] {\cellHaswellLoopsTwoD};
      \addlegendentry{templated, blocking $4\times 3$};
      \addplot table[x={deg}, y={UnrollFive}] {\cellHaswellLoopsTwoD};
      \addlegendentry{templated, blocking $5\times 2$};
      \addplot table[x={deg}, y={UnrollEight}] {\cellHaswellLoopsTwoD};
      \addlegendentry{templated, blocking $8\times 1$};
      \addplot table[x={deg}, y={BasicLoop}] {\cellHaswellLoopsTwoD};
      \addlegendentry{templated loop bounds};
      \addplot table[x={deg}, y={NoTemplate}] {\cellHaswellLoopsTwoD};
      \addlegendentry{non-templated loops};
    \end{semilogyaxis}
  \end{tikzpicture}
  \
  \begin{tikzpicture}
    \begin{semilogyaxis}[
      width=0.51\textwidth,
      height=0.42\textwidth,
      xlabel={Polynomial degree $p$},
      tick label style={font=\scriptsize},
      label style={font=\scriptsize},
      xmin=0,xmax=25,
      ymin=1e8,ymax=1e10,
      title style={at={(1,0.952)},anchor=north east,draw=black,fill=white,font=\scriptsize},
      title={3D Laplacian},
      grid,
      cycle list name=colorGPL,
      mark size=1.8,
      semithick
      ]
      \addplot table[x={deg}, y={EvenOdd}] {\cellHaswellLoopsThreeD};
      \addplot table[x={deg}, y={UnrollFour}] {\cellHaswellLoopsThreeD};
      \addplot table[x={deg}, y={UnrollFive}] {\cellHaswellLoopsThreeD};
      \addplot table[x={deg}, y={UnrollEight}] {\cellHaswellLoopsThreeD};
      \addplot table[x={deg}, y={BasicLoop}] {\cellHaswellLoopsThreeD};
      \addplot table[x={deg}, y={NoTemplate}] {\cellHaswellLoopsThreeD};
    \end{semilogyaxis}
  \end{tikzpicture}
  \\
  \ref{legend2DHasw}
  \caption{Analysis of implementation of matrix multiplication kernels on
    throughput of local Laplace cell kernels on fully populated $2\times 8$
    cores Intel Haswell (Xeon E5-2630 v3 @ 2.4 GHz) in 2D and 3D. The standard
    matrix-matrix multiplication loops are provided in four variants, three of
    which use register blocking with $r\times s$ accumulators, running through
    $r$ local shape value rows and $s$ layers of shape values.}
\label{fig:cell_ops_noload}
\end{figure}
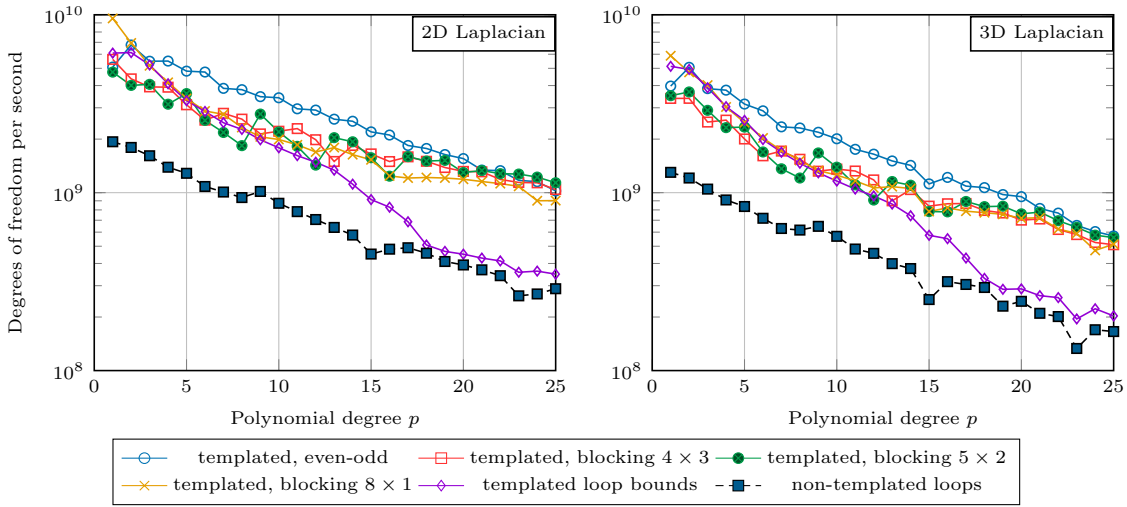

When it comes to the actual implementation of the small matrix-matrix product loop kernels, several optimizations steps beyond a basic loop implementation that passes the loop bounds as run time variables are selected in our implementation. Fig.~\ref{fig:cell_ops_noload} compares these
optimizations in terms of the number of degrees of freedom
processed per second when working on all cores of the $2\times 8$ Intel
Haswell system. The vectorization over elements as presented in the previous
subsection is used, without the tensor-product tiling to ease the
implementation of register blocking schemes.
\begin{itemize}
\item \textbf{Template parameter on loop bounds.} This optimization is
  essential for the short loops at small polynomial degrees, as it allows the
  compiler to completely unroll the loops and
  to re-arrange operations to improve instruction flow.
  Fig.~\ref{fig:cell_ops_noload} shows that performance increases by
  a factor of three approximately. We found best
  performance when keeping the data input along a one-dimensional line of the
  data array in registers and loading the entries in the 1D interpolation
  matrix from L1 cache of full vectorization width.
\item For higher degrees, the compiler's heuristics do not generate optimal
  matrix multiplication code from the templated loops alone. Throughput is
  considerably improved by \textbf{loop unrolling and register blocking} as classically used in
  state-of-the-art matrix-matrix multiplication \texttt{gemm} kernels and
  appearing in \texttt{libxsmm} \cite{libxsmm17}. For the reported results, we manually
  apply $4\times 3$, $5\times 2$, and $8\times 1$ unrolling with 12, 10 and 8
  independent accumulators, with the first number referring to the blocking
  within the coefficient matrix and the second number the lines along which
  blocking is applied. At low degrees where not enough data streams are
  available for blocking, appropriate remainder code is generated. The $5\times 2$ blocked variant reaches up to 500
  GFLOP/s in 2D (75\% of the theoretical peak of 666 GFLOP/s at maximal AVX
  turbo) for degrees between 13 and 25 and up to 420 GFLOP/s for degrees 9,
  14, and 19 in 3D (60\% of arithmetic peak).
\item \textbf{Even-odd decomposition of local matrix-vector kernels.} For the
  case that integration points are symmetric, shape functions symmetric, and derivatives skew-symmetric with
  respect to the center of the cell, there are
  only $k^2/2$ unique entries of the $k^2$ total entries in the 1D
  interpolation or derivative matrices. Thus, working separately on the even
  and odd components of the vector \cite[Sec.~3.5.3]{Kopriva09} reduces the
  operation count for a one-dimensional kernel from $k(2k-1)$ arithmetic
  operations ($k$ multiplications, $k(k-1)$ fused multiply-add operations, FMAs) to
  \begin{equation}\label{eq:complexity_evenodd}
    2k \text{ additions/subtractions, } k \text{ multiplications, and } \lfloor k(k-2)/2\rfloor \text{ FMAs,}
  \end{equation}
  where the additions and subtractions are spent on adding and subtracting the
  first and last vector entries, the second and the second to last, and so on,
  to extract the even and odd parts of the vector.  Besides reducing
  arithmetics, the even-odd decomposition also provides for a natural
  $2\times 1$ loop unrolling. The experiments show a significantly
  higher performance that is more regular than the various loop blocking
  approaches with the full matrix. As reported in Fig.~\ref{fig:cell_ops_arch}
  below, the GFLOP/s go down to around 350 GFLOP/s as compared to the full
  matrix multiplication, but the GFLOP/s rates are secondary quantities
  anyway. At very high degrees, $p>20$, the even-odd implementation could be
  enhanced by further blocking.
\end{itemize}

\subsection{Compute performance on CPUs and Xeon Phi}

We now analyze the throughput of the best kernel for the cell Laplacian, invoking
$2d$ tensor product kernels in two \texttt{basis\_change} calls and $2d$
tensor product kernels in two \texttt{collocation\_derivative} calls according to Table~\ref{tab:tensor_costs}, each
implemented with even-odd decomposition with loop tiling and templated loop bounds, on the three
Intel processors presented in Table~\ref{tab:systems}.

\begin{figure}
  \centering
  \begin{tikzpicture}
    \begin{semilogyaxis}[
      width=0.49\textwidth,
      height=0.48\textwidth,
      xlabel={Polynomial degree $p$},
      ylabel={Degrees of freedom per second},
      tick label style={font=\scriptsize},
      label style={font=\scriptsize},
      legend to name=legendThroughput,
      legend columns=4,
      legend style={font=\scriptsize},
      xmin=0,xmax=25,
      ymin=5e8,ymax=3.2e10,
      grid,
      cycle list name=colorGPL,
      mark size=1.8,
      semithick
      ]
      \addplot table[x={deg}, y={HaswellTwo}] {\cellLoopsArchitecturesTime};
      \addlegendentry{2D, Haswell};
      \addplot table[x={deg}, y={BroadwellTwo}] {\cellLoopsArchitecturesTime};
      \addlegendentry{2D, Broadwell};
      \addplot table[x={deg}, y={KnlTwo}] {\cellLoopsArchitecturesTime};
      \addlegendentry{2D, KNL 64 threads};
      \addplot[gnuplot@green,densely dotted] table[x={degree}, y={LaplaceManu}] {\cellKnlVectEOTwo};
      \addlegendentry{2D, KNL 128 threads};
      \pgfplotsset{cycle list shift=-1};
      \addplot table[x={deg}, y={HaswellThree}] {\cellLoopsArchitecturesTime};
      \addlegendentry{3D, Haswell};
      \addplot table[x={deg}, y={BroadwellThree}] {\cellLoopsArchitecturesTime};
      \addlegendentry{3D, Broadwell};
      \addplot[black,every mark/.append style={fill=gnuplot@darkblue!80!black},mark=triangle*] table[x={deg}, y={KnlThree}] {\cellLoopsArchitecturesTime};
      \addlegendentry{3D, KNL 64 threads};
      \addplot[black,densely dashdotted] table[x={degree}, y={LaplaceManu}] {\cellKnlVectEOThree};
      \addlegendentry{3D, KNL 128 threads};
    \end{semilogyaxis}
  \end{tikzpicture}
  \
  \begin{tikzpicture}
    \begin{axis}[
      width=0.49\textwidth,
      height=0.48\textwidth,
      xlabel={Polynomial degree $p$},
      ylabel={GFLOP/s},
      tick label style={font=\scriptsize},
      label style={font=\scriptsize},
      xmin=0,xmax=25,
      ymin=0,ymax=1200,
      grid,
      cycle list name=colorGPL,
      mark size=1.8,
      semithick
      ]
      \addplot table[x={deg}, y={HaswellTwo}] {\cellLoopsArchitecturesGflop};
      \addplot table[x={deg}, y={BroadwellTwo}] {\cellLoopsArchitecturesGflop};
      \addplot table[x={deg}, y={KnlTwo}] {\cellLoopsArchitecturesGflop};
      \addplot[gnuplot@green,densely dotted] table[x={degree}, y={KnlLaplM}] {\cellGflopTwo};
      \pgfplotsset{cycle list shift=-1};
      \addplot table[x={deg}, y={HaswellThree}] {\cellLoopsArchitecturesGflop};
      \addplot table[x={deg}, y={BroadwellThree}] {\cellLoopsArchitecturesGflop};
      \addplot[black,every mark/.append style={fill=gnuplot@darkblue!80!black},mark=triangle*] table[x={deg}, y={KnlThree}] {\cellLoopsArchitecturesGflop};
      \addplot[black,densely dashdotted] table[x={degree}, y={KnlLaplM}] {\cellGflopThree};
    \end{axis}
  \end{tikzpicture}
  \\
  \ref{legendThroughput}
  \caption{Throughput of local cell kernels on $2\times 8$ core Intel Xeon
    E5-2630 v3 (Haswell), $2\times 14$ core Intel Xeon E5-2690 v4 (Broadwell),
    and 64 core Intel Xeon Phi 7210 (KNL) for cell integrals of
    Laplacian in 2D and 3D.}
\label{fig:cell_ops_arch}
\end{figure}
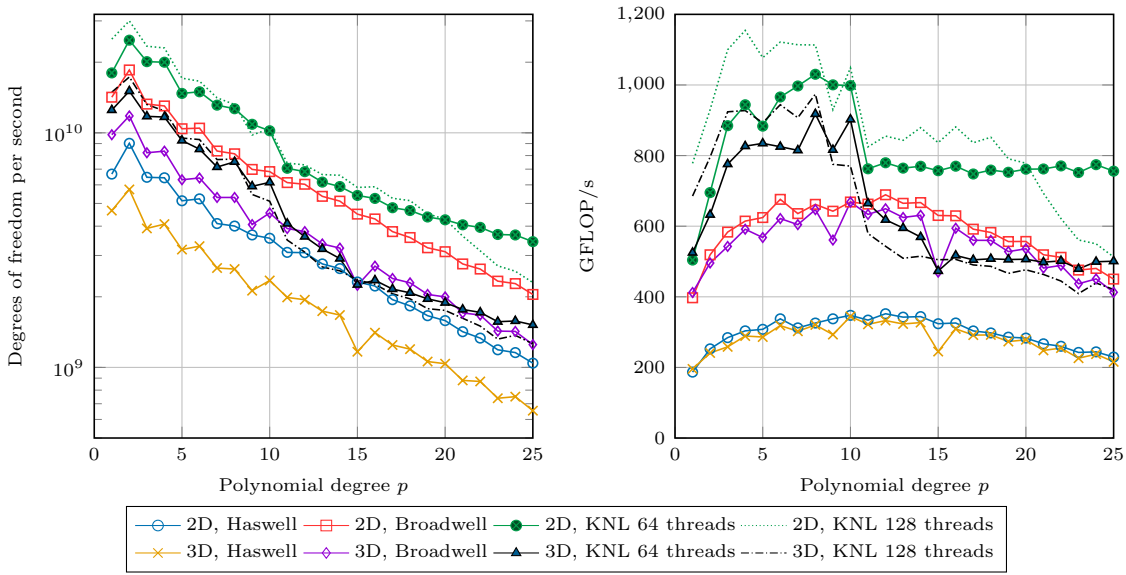

Fig.~\ref{fig:cell_ops_arch} shows that the Haswell system reaches 340--350 GFLOP/s
in both 2D and 3D around $p=10$. The number of floating point
operations is obtained by multiplying the cost per tensor product kernel from
Eq.~\eqref{eq:complexity_evenodd} by $4d$, the number of calls to tensor product kernels
according to Table~\ref{tab:tensor_costs}, and including the cost per quadrature point
on a Cartesian mesh involving $2d+1$ multiplications. We have verified that
these numbers are accurate to within 10\% of the measured arithmetic throughput with
likwid. For the mix of additions, multiplications, and FMAs in the even-odd
kernel according to Eq.~\eqref{eq:complexity_evenodd}, the throughput ceiling
can be computed to be 340 GFLOP/s for $p=1$, 465 GFLOP/s for $p=3$, 550 GFLOP/s for $p=5$ and
approaching 600 GFLOP/s for $p=20$. In other words, the code reaches up to
40--60\% of the possible arithmetic performance.\footnote{Performance is
  sensitive to the compiler's choices in register allocation and loop
  unrolling. For example, we recorded \texttt{g++} of version 7.1 to be 10\%
  slower on $p=8$ than \texttt{g++} version 6.3, but almost equal for many
  other degrees.}  Also, the actual performance limit is often the L1 cache read
and write access rather than pure arithmetics. On Broadwell, the best performance
is 690 GFLOP/s in 2D and 670 GFLOP/s in 3D, again up to 60\% of the possible
arithmetic throughput assuming \emph{highest AVX turbo} for the actual upper
range of between 700 and 1150 GFLOP/s. As expected, we reach similar
percentages of the arithmetic peak on both Haswell and Broadwell
because the test case is compute only. Given that our code reaches more than
50\% of FMA peak with the even-odd decomposition, it outperforms any version of
straight matrix multiplication as documented in Fig.~\ref{fig:cell_ops_noload}.
Similar GFLOP/s rates are achieved in the mass
matrix application (at twice the DoF/s throughput) and the advection operation
(at 1.5 times the DoF/s throughput).

On the KNL many-core processor, the gap to the theoretical value of 2250
GFLOP/s is larger. In 3D, a decrease in arithmetic performance is
observed for $p\approx 11$ where the inner kernel exhausts the 512 kiB of L2
cache per core and an increasingly larger part of the local integration data
needs to be fetched from the MCDRAM memory. Note that performance for $p>10$
is only around 50 GFLOP/s when binding the kernel to the slow DDR4 RAM with
\texttt{numactl --membind=0} rather than the fast 16 GB of MCDRAM
\cite{Jeffers16}. The tensor product tiling of Algorithm
\ref{alg:kernel_tiling} and the generated machine
code are more important on KNL than on the CPUs and in particular once the L2 cache is
exceeded. For example, the jump between degrees 10 and 11 in 2D is due to machine code
generation effects, with the whole 1D input line of sum factorization fitting
into registers for $p\leq 10$ and re-loading them for higher
degrees. Furthermore, we observe considerably lower GFLOP/s rates in 3D than in 2D, as opposed to the Haswell and Broadwell systems, which could be explained by more loads and stores from caches as compared to arithmetics in 3D. Nonetheless, our results show that one can reach around 1 TFLOP/s
for low to moderate polynomial degrees or 50\% of peak arithmetic
performance, which is extremely good given the restricted hardware features of
the KNL microarchitecture. Note that for low degrees $p\leq 7$ in 3D and for
$p\leq 20$ in 2D, KNL with two-way simultaneous multithreading (128 threads)
performs better than on 64 threads, but the picture is reversed for higher
degrees when too much local data is in flight.

\subsection{Compute throughput of cell and face integrals}\label{sec:cells_and_faces}

Finally, we include face integrals in the comparison.  In order to
efficiently evaluate face integrals, we use a nodal basis for advection with
the nodes of the $k$-point Gauss--Lobatto--Legendre quadrature formula, i.e.,
one node placed at each of the 1D interval endpoints. This allows to directly
pick the $k^{d-1}$ node values on the face without interpolation normal to the
face over all $k^d$ points. Likewise, a Hermite-like basis
is chosen for the Laplacian \eqref{eq:poisson_sip} that can compute the
solution's value and derivative on the face from $2k^{d-1}$ solution values on
cubic and higher degree polynomials, $k\geq 4$, according to the
\texttt{face\_normal\_interpolation} kernel from
Sec.~\ref{sec:sum_factorization}. Gaussian quadrature on $k^d$ points is used for both cases, necessitating use of \texttt{basis\_change} in addition to the derivative kernel. Note that the face integrals of the
Laplacian involve more than twice as much arithmetics as compared to the
advection operator, see the second column of Table~\ref{tab:tensor_costs}.

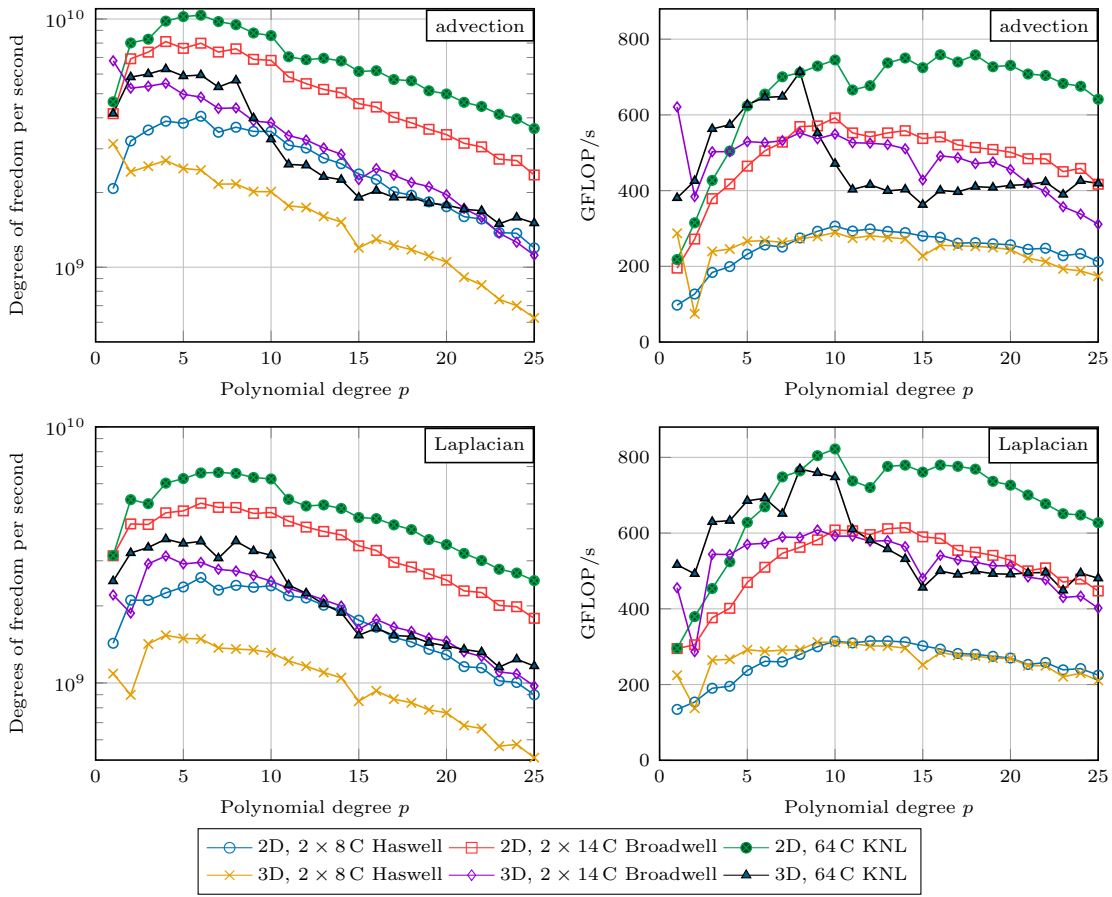
\begin{figure}
  \centering
  \begin{tikzpicture}
    \begin{semilogyaxis}[
      width=0.49\textwidth,
      height=0.4\textwidth,
      xlabel={Polynomial degree $p$},
      ylabel={Degrees of freedom per second},
      tick label style={font=\scriptsize},
      label style={font=\scriptsize},
      legend to name=legendThroughputFace,
      legend columns=3,
      legend style={font=\scriptsize},
      title style={at={(1,0.95)},anchor=north east,draw=black,fill=white,font=\scriptsize},
      title={advection},
      xmin=0,xmax=25,
      ymin=5e8,ymax=1.1e10,
      grid,
      cycle list name=colorGPL,
      mark size=1.8,
      semithick
      ]
      \addplot table[x={deg}, y={HswConvect}] {\dgDofsTwo};
      \addlegendentry{2D, $2\times 8$\,C Haswell};
      \addplot table[x={deg}, y={BdwConvect}] {\dgDofsTwo};
      \addlegendentry{2D, $2\times 14$\,C Broadwell};
      \addplot table[x={deg}, y={KnlConvect}] {\dgDofsTwo};
      \addlegendentry{2D, $64$\,C KNL};
      \addplot table[x={deg}, y={HswConvect}] {\dgDofsThree};
      \addlegendentry{3D, $2\times 8$\,C Haswell};
      \addplot table[x={deg}, y={BdwConvect}] {\dgDofsThree};
      \addlegendentry{3D, $2\times 14$\,C Broadwell};
      \addplot[black,every mark/.append style={fill=gnuplot@darkblue!80!black},mark=triangle*] table[x={deg}, y={KnlConvect}] {\dgDofsThree};
      \addlegendentry{3D, $64$\,C KNL};
    \end{semilogyaxis}
  \end{tikzpicture}
  \
  \begin{tikzpicture}
    \begin{axis}[
      width=0.49\textwidth,
      height=0.4\textwidth,
      xlabel={Polynomial degree $p$},
      ylabel={GFLOP/s},
      tick label style={font=\scriptsize},
      label style={font=\scriptsize},
      xmin=0,xmax=25,
      ymin=0,ymax=880,
      title style={at={(1,0.95)},anchor=north east,draw=black,fill=white,font=\scriptsize},
      title={advection},
      grid,
      cycle list name=colorGPL,
      mark size=1.8,
      semithick
      ]
      \addplot table[x={deg}, y={HswConvect}] {\dgGflopTwo};
      \addplot table[x={deg}, y={BdwConvect}] {\dgGflopTwo};
      \addplot table[x={deg}, y={KnlConvect}] {\dgGflopTwo};
      \addplot table[x={deg}, y={HswConvect}] {\dgGflopThree};
      \addplot table[x={deg}, y={BdwConvect}] {\dgGflopThree};
      \addplot[black,every mark/.append style={fill=gnuplot@darkblue!80!black},mark=triangle*] table[x={deg}, y={KnlConvect}] {\dgGflopThree};
    \end{axis}
  \end{tikzpicture}
  \\
  \begin{tikzpicture}
    \begin{semilogyaxis}[
      width=0.49\textwidth,
      height=0.4\textwidth,
      xlabel={Polynomial degree $p$},
      ylabel={Degrees of freedom per second},
      tick label style={font=\scriptsize},
      label style={font=\scriptsize},
      title style={at={(1,0.95)},anchor=north east,draw=black,fill=white,font=\scriptsize},
      title={Laplacian},
      xmin=0,xmax=25,
      ymin=5e8,ymax=1e10,
      grid,
      cycle list name=colorGPL,
      mark size=1.8,
      semithick
      ]
      \addplot table[x={deg}, y={HswLaplace}] {\dgDofsTwo};
      \addplot table[x={deg}, y={BdwLaplace}] {\dgDofsTwo};
      \addplot table[x={deg}, y={KnlLaplace}] {\dgDofsTwo};
      \addplot table[x={deg}, y={HswLaplace}] {\dgDofsThree};
      \addplot table[x={deg}, y={BdwLaplace}] {\dgDofsThree};
      \addplot[black,every mark/.append style={fill=gnuplot@darkblue!80!black},mark=triangle*] table[x={deg}, y={KnlLaplace}] {\dgDofsThree};
    \end{semilogyaxis}
  \end{tikzpicture}
  \
  \begin{tikzpicture}
    \begin{axis}[
      width=0.49\textwidth,
      height=0.4\textwidth,
      xlabel={Polynomial degree $p$},
      ylabel={GFLOP/s},
      tick label style={font=\scriptsize},
      label style={font=\scriptsize},
      xmin=0,xmax=25,
      ymin=0,ymax=880,
      title style={at={(1,0.95)},anchor=north east,draw=black,fill=white,font=\scriptsize},
      title={Laplacian},
      grid,
      cycle list name=colorGPL,
      mark size=1.8,
      semithick
      ]
      \addplot table[x={deg}, y={HswLaplace}] {\dgGflopTwo};
      \addplot table[x={deg}, y={BdwLaplace}] {\dgGflopTwo};
      \addplot table[x={deg}, y={KnlLaplace}] {\dgGflopTwo};
      \addplot table[x={deg}, y={HswLaplace}] {\dgGflopThree};
      \addplot table[x={deg}, y={BdwLaplace}] {\dgGflopThree};
      \addplot[black,every mark/.append style={fill=gnuplot@darkblue!80!black},mark=triangle*] table[x={deg}, y={KnlLaplace}] {\dgGflopThree};
    \end{axis}
  \end{tikzpicture}
  \\
  \ref{legendThroughputFace}
  \caption{Throughput of cell and face kernels for advection
    and Laplacian in 2D and 3D
    without vector access on $2\times 8$ core Intel Xeon E5-2630 v3
    (Haswell), $2\times 14$ core Intel Xeon E5-2690 v4 (Broadwell) and 64 core
    Intel Xeon Phi 7210 (KNL).}
\label{fig:dg_ops_arch}
\end{figure}

Fig.~\ref{fig:dg_ops_arch} presents the throughput of the combined cell and face integrals on the three hardware
systems of Table~\ref{tab:systems}. In order to avoid the
indirect addressing associated with face integrals in general, we have
considered the artificial case of periodic boundary conditions within
the same cell that can directly use the vectorized data storage according to
Fig.~\ref{fig:vectorization_over_cells}. The results in
Fig.~\ref{fig:dg_ops_arch} confirm the performance results from the cell
integrals in the previous figures, with up to 620 GFLOP/s on the Broadwell
system and 820 GFLOP/s on KNL. Note that face integrals contribute with more than two thirds of the
arithmetic work of the Laplacian for polynomial degrees up to five. For the 3D
Laplacian, our code reaches a throughput of more than $3\cdot 10^9$ degrees of
freedom per second on KNL for $2\leq p \leq 10$ and more than $2.6 \cdot 10^9$
degrees of freedom on Broadwell for $3\leq p \leq 9$.

\section{Data access patterns and MPI parallelization}\label{sec:vector_access}

In this section, we analyze the performance of the full operator evaluation including the actual data access
patterns of DG cell and face integrals into the input and output vectors, as well as
parallelization.

To exploit the parallelism of multi-core processors that are connected by
high-speed networks in modern petascale machines, two different
parallelization concepts are commonly used, the shared-memory paradigm of
OpenMP/pthreads and the distributed memory concept implemented by the message passing
interface (MPI). Increasingly, a mixture of both is applied to exploit intra-node
and internode parallelism, respectively.
MPI parallelism for finite-element based computations usually relies on
domain decomposition that splits the cells among the
processors. Each processor only works on its portion of the mesh. In order
to exchange information on the border between subdomains,
ghost elements around the locally owned subdomain are used. In our implementation, we
assume one layer of ghost elements around the owned
cells to be present, supported by the massively parallel algorithm from
\cite{bbhk11}. The particular form of the mesh partitioning is immaterial, as
long as the information provided by the mesh infrastructure allows for a
unique identification of the degrees of freedom in the locally owned cells and
the ghost layer.

For shared-memory parallelism, loops over the mesh entities are
split across the participating threads. Some coordination is necessary to
avoid race conditions when integrals from several faces go to the same vector
entries. A common scheme to avoid race conditions in many DG codes is to
create a temporary storage that holds independent data for all the faces. This
involves an initial loop over the cells where the face data is collected (and
cell integrals are computed), a loop over the faces where only the separate
face storage is referenced, and a final loop over the cells that collects the
face integrals and lifts them onto the cells.
The face-normal part of interpolation
and integration steps of face integrals are processed in the first and third
steps, whereas operations within faces are done in the face loop, see also the
algorithm layout described extensively in \cite{hw08}. Such a strategy
involves additional data transfer since the cells are accessed twice and
separate global face storage is involved. Even though it would be conceivable
to keep the data storage lower with dynamic dependency-based task scheduling,
the authors' experience from \cite{Kormann11,Kronbichler17} suggests that available
implementations such as Intel Threading Building Blocks \cite{Rei07} or OpenMP tasks do
typically lead to significant memory access from remote NUMA domains and other
cache or prefetcher inefficiencies that lower application performance once using 10
or more cores. In this work, we therefore do not consider the various
possibilities of shared-memory parallelization and analyze MPI for
parallelization within the shared memory of a node.

The design pattern of using \textbf{face-separate storage} is also common in
MPI-only codes, see e.g.~\cite{Hindenlang12}, so we consider that storage
scheme implemented with MPI communication in our analysis below to
contrast this very common DG implementation technique against the proposed
methods. To highlight the properties of separate loops for cell and face
integrals, respectively, we start by an analysis of the cell integral only,
which similarly translate to applying DG mass matrices or inverse mass
matrices. We consider large vector sizes of around 10--50 million degrees of
freedom to eliminate any cache effects.

\subsection{Vector access analysis for cell integrals only}

\begin{figure}
  \centering
  \begin{tikzpicture}
    \begin{semilogyaxis}[
      width=0.49\textwidth,
      height=0.4\textwidth,
      xlabel={Polynomial degree $p$},
      ylabel={Degrees of freedom per second},
      tick label style={font=\scriptsize},
      label style={font=\scriptsize},
      legend to name=legendThroughputAccess,
      legend columns=3,
      legend style={font=\scriptsize},
      title style={at={(1,0.95)},anchor=north east,draw=black,fill=white,font=\scriptsize},
      title={2D},
      xmin=0,xmax=25,
      ymin=5e8,ymax=3.1e10,
      grid,
      cycle list name=colorGPL,
      mark size=1.8,
      semithick
      ]
      \addplot table[x={degree}, y={HswLaplace}] {\cellWithAccessTwo};
      \addlegendentry{$2\times 8$\,C Haswell};
      \addplot table[x={degree}, y={BdwLaplace}] {\cellWithAccessTwo};
      \addlegendentry{$2\times 14$\,C Broadwell};
      \addplot table[x={degree}, y={KnlLaplace}] {\cellWithAccessTwo};
      \addlegendentry{$64$\,C KNL};
      \addplot[gnuplot@darkblue] table[x={degree}, y={LaplaceManu}] {\cellHaswellVectEOTwo};
      \addplot[red!80!white] table[x={degree}, y={LaplaceManu}] {\cellBroadwellVectEOTwo};
      \addplot[gnuplot@green] table[x={degree}, y={LaplaceManu}] {\cellKnlVectEOTwo};
      \addplot[gnuplot@darkblue,densely dashed] coordinates {
        (0,3.96e+9) (25,3.96e+9)
      };
      \addplot[red!80!white,densely dashed] coordinates {
        (0,4.67e+9) (25,4.67e+9)
      };
      \node[red!80,outer sep=0.5pt,inner sep=0.5pt] at (axis cs:21.5,4.67e+9) [anchor=south,fill=white] {\scriptsize 112 GB/s};
      \addplot[gnuplot@green,densely dashed] coordinates {
        (0,1.87e+10) (25,1.87e+10)
      };
      \node[gnuplot@green,outer sep=0.5pt,inner sep=0.5pt] at (axis cs:21.5,1.87e+10) [anchor=north,fill=white] {\scriptsize 450 GB/s};
    \end{semilogyaxis}
  \end{tikzpicture}
  \
  \begin{tikzpicture}
    \begin{semilogyaxis}[
      width=0.49\textwidth,
      height=0.4\textwidth,
      xlabel={Polynomial degree $p$},
      ylabel={Degrees of freedom per second},
      tick label style={font=\scriptsize},
      label style={font=\scriptsize},
      title style={at={(1,0.95)},anchor=north east,draw=black,fill=white,font=\scriptsize},
      title={3D},
      xmin=0,xmax=25,
      ymin=5e8,ymax=3.1e10,
      grid,
      cycle list name=colorGPL,
      mark size=1.8,
      semithick
      ]
      \addplot table[x={degree}, y={HswLaplace}] {\cellWithAccessThree};
      \addplot table[x={degree}, y={BdwLaplace}] {\cellWithAccessThree};
      \addplot table[x={degree}, y={KnlLaplace}] {\cellWithAccessThree};
      \addplot[gnuplot@darkblue] table[x={degree}, y={LaplaceManu}] {\cellHaswellVectEOThree};
      \addplot[red!80!white] table[x={degree}, y={LaplaceManu}] {\cellBroadwellVectEOThree};
      \addplot[gnuplot@green] table[x={degree}, y={LaplaceManu}] {\cellKnlVectEOThree};
      \addplot[gnuplot@darkblue,densely dashed] coordinates {
        (0,3.96e+9) (25,3.96e+9)
      };
      \node[gnuplot@darkblue,outer sep=1pt,inner sep=1pt] at (axis cs:21,3.96e+9) [anchor=north,fill=white] {\scriptsize 95 GB/s};
      \addplot[red!80!white,densely dashed] coordinates {
        (0,4.67e+9) (25,4.67e+9)
      };
      \node[red!80,outer sep=1pt,inner sep=1pt] at (axis cs:21,4.67e+9) [anchor=south,fill=white] {\scriptsize 112 GB/s};
      \addplot[gnuplot@green,densely dashed] coordinates {
        (0,1.87e+10) (25,1.87e+10)
      };
      \node[gnuplot@green,outer sep=1pt,inner sep=1pt] at (axis cs:21,1.87e+10) [anchor=north,fill=white] {\scriptsize 450 GB/s};
    \end{semilogyaxis}
  \end{tikzpicture}
  \\
  \ref{legendThroughputAccess}
  \caption{Verification of performance model for cell integrals of Laplacian
    in 2D and 3D modeling the global vector access on
    $2\times 8$ core Intel Xeon E5-2630 v3 (Haswell), $2\times 14$ core Intel
    Xeon E5-2690 v4 (Broadwell) and 64 core 2nd generation Intel Xeon Phi 7210
    (KNL) (using hyperthreading with 128 processes). The solid lines indicate
    the computational throughput according to Fig.~\ref{fig:cell_ops_arch} and
    the dashed lines the memory bandwidth limit of a stream triad kernel,
    assuming two vector reads and one vector write.}
\label{fig:cell_access_arch}
\end{figure}
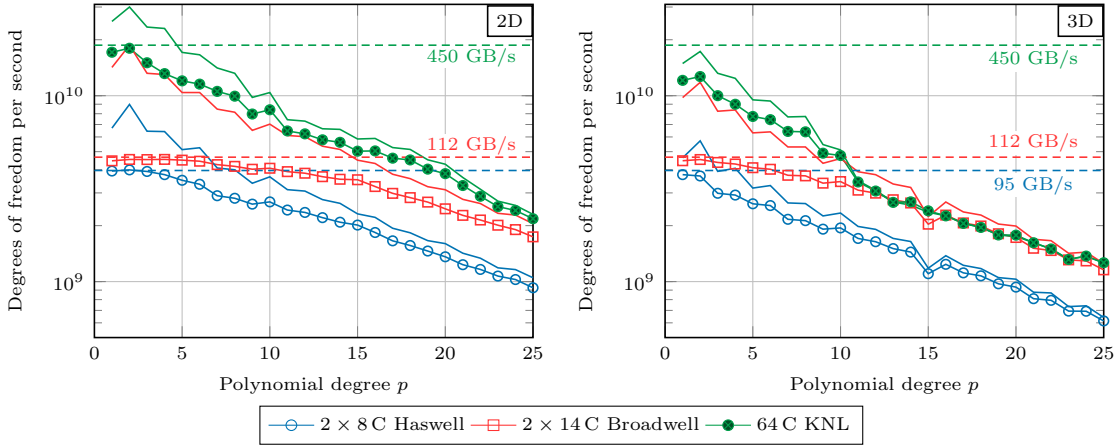

In Fig.~\ref{fig:cell_access_arch}, we consider the cell benchmark from
Fig.~\ref{fig:cell_ops_arch} including the access to the global solution
vectors. We assume a Cartesian geometry where our implementation simply uses
the same inverse Jacobian $\mathcal J^{(e)}$ matrix for all the quadrature
points, not accessing additional global memory. Thus, the main memory transfer
in this algorithm is one vector read for the input vector, one vector write
for the output vector, and one vector read operation on the output vector
either because of an accumulate-into pattern or due to the read-for-ownership
memory access pattern \cite{Hager11}. The results in
Fig.~\ref{fig:cell_access_arch} are contrasted against the two theoretical
performance limits, the memory bandwidth limit of the vector access as
measured by a STREAM triad test which give 95 GB/s, 112 GB/s, and 450 GB/s for
the Haswell, Broadwell, and KNL systems, respectively, and the arithmetic
throughput from the previous section. On a test similar to the one described
in Fig.~\ref{fig:cell_memory_access}, we confirmed with the likwid tool that
data transferred through the memory hierarchy has not changed except for an
additional access of 24 bytes per degree of freedom that pass the solution
vectors through all cache levels.

The results in Fig.~\ref{fig:cell_access_arch} suggest that the memory
bandwidth is the limit to throughput in 2D and at low and medium polynomial
degrees $k\leq 10$, in particular for architectures with more cores such as
Broadwell. On the other hand, on the Haswell and KNL systems where the memory
bandwidth is higher in relation, the code is mostly compute limited in 3D,
despite the high level of optimization presented above. In both cases, the
envelope established by the memory throughput and the arithmetic throughput
closely describes the achieved performance. The remaining gap at
intermediate degrees is mostly explained by effects not captured by the simple
distinction between memory bandwidth limit and arithmetic peak, but
could be explained by more advanced performance models such as the execution cache memory
model \cite{Hager11}. These experiments lead to the following design choice.

\begin{Requirement}
  Since cell integrals tend to be memory bound already for simple Cartesian meshes,
  face integrals must be interleaved with cell integrals in
  Algorithm~\ref{alg:prototype_dof_op} to re-use vector data that resides in
  caches. For curved geometries that involve significantly lower FLOP/byte
  ratios, a reduction in memory access for vector entries improves performance also on machines
  with abundant memory bandwidth such as KNL with its high-bandwidth memory.
\end{Requirement}

  We emphasize that the compute optimizations presented in
  Sec.~\ref{sec:kernel_local} are indeed essential: the upper memory limit of
  two loads and one store of global vectors is at around
  $4.7 \cdot 10^9$ degrees of freedom per second on 28 cores of Broadwell, which is higher
  than the compute throughput of around $2.8 \cdot 10^9$ degrees of freedom
  per second for the Laplacian in 3D and only slightly below the maximal throughput
  of $5 \cdot 10^9$ for advection in 3D according to Fig.~\ref{fig:dg_ops_arch}. Also, the fact that the Broadwell system tends to be memory bound (for $k<10$) on cell integrals is a result of the high level of optimization of the compute kernels.

\subsection{Vectorization layout for face integrals}

In this subsection, we present data structures to organize vectorization
over several faces for a finite element mesh beyond the compute kernels
presented in Section~\ref{sec:kernel_local}.  Face integrals contain some
unavoidable dispatch code that is invoked at run time, namely the selection of
different data access routines depending on the local face number with respect
to the cell as well as some shuffling of the data interpolated to quadrature
points for 3D meshes where some faces are not in the standard orientation with
respect to the cell's local coordinate system \cite{Agelek17}. Since face
integration kernels are relatively small, the dispatch overhead reduces in
importance if it is around a batch of faces with the exact same code
path.
The demands of integration kernels from Sec.~\ref{sec:sum_factorization} are satisfied by the following two design choices.

\begin{Requirement}
  In order to restrict vector access to only those entries where $S_\text{f}$
  and $D_\text{f}$ from Eq.~\eqref{eq:face_integral} are non-zero, access to
  the solution vector must know the highest order derivative in the evaluation
  routines to only read or write the necessary vector data alongside with the
  basis type such as \texttt{nodal\_on\_faces} or
  \texttt{hermite\_like\_basis}.
\end{Requirement}

\begin{Requirement}\label{req:arrangement_gradients}
  In order to simplify implementation and re-use 2D kernels, the local
  coordinate system on faces is always set such that reference cell gradients
  touch the $d-1$ tangential directions first and the face-normal direction
  comes last by adjusting the order of components in the geometry tensors rather
  than changing indices of evaluators, see Eq.~\eqref{eq:face_integral}.
\end{Requirement}

Data Structure \ref{alg:face_info} lists a slim way of
storing a pair of faces in case of vectorization. In our C++
implementation, the width of the SIMD array is set by a template argument,
allowing to distinguish the SIMD width of single precision \texttt{float}
and double-precision \texttt{double} or between different computer systems. The identification of the faces in a
setup stage runs through an arbitrary quadrilateral or hexahedral mesh
and first creates a separate \texttt{FaceInfo} object for each face. In a second step, it
looks for faces with the same structure: a comparator checks for
the values of \texttt{interior\_face\_number},
\texttt{exterior\_face\_number}, \texttt{subface\_index}, and
\texttt{face\_orientations} and allows merging the faces into the same batch if they are all
equal. Remarks \ref{remark:subface} and \ref{remark:mix_face_types} show
possibilities to relax this requirement for increasing the utilization of the
SIMD lanes for small problem sizes.

\begin{datastructure}
  \caption{\texttt{struct FaceInfo<SIMD\_WIDTH>}: Face data structure with vectorization}
  \label{alg:face_info}
  \begin{itemize}
  \setlength{\itemsep}{0pt}
  \setlength{\parsep}{0pt}
  \setlength{\parskip}{0pt}
  \item \texttt{unsigned int interior\_cell\_numbers[SIMD\_WIDTH]}: Short array
    of indices to the cells flagged as $e^-$ (interior); holds invalid value
    $2^{32}-1$ if SIMD array not completely filled
  \item \texttt{unsigned int exterior\_cell\_numbers[SIMD\_WIDTH]}: Short array
    of indices to the exterior cells $e^+$; holds invalid value
    $2^{32}-1$ if face at boundary or SIMD array not completely filled
  \item \texttt{unsigned char interior\_face\_number}: local face number
    $l_f^-\in[0,\ldots,2d)$ within cell $e^-$
  \item \texttt{unsigned char exterior\_face\_number}: local face number
    $l_f^+\in[0,\ldots,2d)$ within cell $e^+$ for inner faces; for boundary
    faces, this storage location is used for storing a boundary id to
    distinguish between various types of boundaries in user code
    (e.g. Dirichlet or Neumann)
  \item \texttt{unsigned char subface\_index}: local subface index in
    $[0, 2^{d-1})$ of the exterior side of the face in case of adaptivity ($2:1$ mesh
    ratio required), a number of $2^8-1$ indicates a uniform face
  \item \texttt{unsigned char face\_orientations}: Index of face orientation
    in $[0, 8)$ for non-standard orientation on $-$ side and $[8, 16)$ on $+$
    side
  \end{itemize}
\end{datastructure}

\begin{Remark}\label{remark:subface}
  Faces of different \texttt{subface\_index}, e.g. faces with hanging nodes
  and others without, can be combined into the same SIMD array for
  small problem sizes if the interpolation matrices are created e.g.~by
  selection masks into the respective interpolation array (\texttt{vblend*}
  instructions on x86-64 SIMD).
\end{Remark}

\begin{Remark}\label{remark:mix_face_types}
  For special face interpolation types \texttt{nodal\_at\_face} or
  \texttt{hermite\_type\_basis} in the \texttt{face\_normal\_interpolation} of
  Sec.~\ref{sec:sum_factorization}, gather access can be generalized to allow
  for combining partially filled SIMD lanes of different values of
  \texttt{interior\_face\_number} and
  \texttt{exterior\_face\_number}. However, this increases the amount of
  indirect addressing and is not considered in this work. For this scheme,
  Design Choice \ref{req:arrangement_gradients} is a prerequisite to keep the
  geometry application vectorized without further cross-lane permutations.
\end{Remark}

\subsection{Numbering of degrees of freedom and vector access}

In DG, all degrees of freedom of a cell can be identified by a single index
because there are no hard continuity constraints linking the indices to the ones on neighboring cells. Nonetheless, the details of
index storage are important for reaching high throughput. Data structure
\ref{alg:dof_info} details how to realize an efficient scheme
in a generic finite element code.
In Section~\ref{sec:kernel_local}, we have proposed an interleaved numbering of the degrees
of freedom for $n_\text{lanes}$ cells according to
Fig.~\ref{fig:vectorization_over_cells} as the main option. Thus, explicit
vector read operations and data permutations can be skipped in favor of
passing a pointer to vector data, e.g. \texttt{\_\_m256d}, to the sum
factorization kernels for cell integrals. However, the access for the cell data for
face integrals must handle several different cases. As an example, consider
the advection equation \eqref{eq:advection} on the degrees of freedom laid out
in Fig.~\ref{fig:vectorization_over_cells}. Let us assume we work on the faces
$f_1=e_1\cap e_2$, $f_2=e_2\cap e_5$, $f_3=e_3\cap e_4$, $f_4=e_4\cap
e_6$, where we have
\begin{equation*}
  \texttt{FaceInfo<4>} =
 \left\{\begin{array}{ll} \texttt{interior\_cell\_numbers} & \{1, 2, 3, 4\}\\
          \texttt{exterior\_cell\_numbers} & \{2, 5, 4, 7\} \\
          \texttt{interior\_face\_number} & 1\\
          \texttt{exterior\_face\_number} & 0\\
          \texttt{subface\_index} & 255\\
          \texttt{face\_orientation} & 0
        \end{array}
      \right\},
\end{equation*}
assuming that local face numbers are 0 for the left face of a
cell and 1 for the right face of a cell.
In this case,
the interior cell numbers $e^-$ of the faces coincide with the
interleaved numbering of the cell, allowing direct packed access. Hence, the face is identified as \texttt{IndexStorageVariants::\-interleaved\_contiguous} for the interior side $e^-$.
For the cell data
related to the exterior side $e^+$, however, the index numbering is not
contiguous, as the indices to be accessed are $\{1, 64, 3, 66,
\ldots\}$. Thus, for the \texttt{IndexStorageVariants::interleaved\_contiguous\_strided} format a gather/scatter operation is needed to access the vector
entries. When passing
from the first four to the next four indices, a constant stride of
4 (SIMD width) appears so only the base address of the gather operation
needs to be recomputed.

\begin{wrapfigure}{R}{0.31\textwidth}
  \begin{tikzpicture}
    \fill[fill=yellow!50!white] (0,-1.2) rectangle (2.5,0);
    \fill[fill=red!20!white] (0,1.2) rectangle (2.5,0);
    \fill[fill=blue!50!green!20!white] (2.5,-1.2) rectangle (4.3,1.2);
    \draw(0,0) -- (4.3,0);
    \draw(0,1) -- (4.3,1);
    \draw(0,-1) -- (4.3,-1);
    \draw(0.5,-1.2) -- (0.5,1.2);
    \draw(1.5,-1.2) -- (1.5,1.2);
    \draw(2.5,-1.2) -- (2.5,1.2);
    \draw(3.5,-1.2) -- (3.5,1.2);
    \draw(1,-0.9) node[fill=white,fill opacity=0.8]  {\scriptsize processor $p_1$};
    \draw (1,0.9) node[fill=white,fill opacity=0.8] {\scriptsize processor $p_2$};
    \draw (3.45,0) node[fill=white,fill opacity=0.8] {\scriptsize processor $p_3$};
    \node at (1,-0.5) {\scriptsize $e_1$};
    \node at (2,-0.5) {\scriptsize $e_2$};
    \node at (1,0.5) {\scriptsize $e_3$};
    \node at (2,0.5) {\scriptsize $e_4$};
    \node at (3,-0.5) {\scriptsize $e_5$};
    \node at (4,-0.5) {\scriptsize $e_6$};
    \node at (3,0.5) {\scriptsize $e_7$};
    \node at (4,0.5) {\scriptsize $e_8$};
  \end{tikzpicture}
  \caption{Illustration of a situation where three processors meet at a corner
    and at least one processor must resort to a mixed-stride case within the
    index storage.}
  \label{fig:ghost-mixed-stride}
\end{wrapfigure}
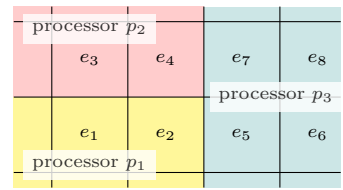

Data structure \ref{alg:dof_info} lists
\texttt{IndexStorageVariants::inter\-leaved\_contiguous\_mixed\_strides} as a
third interleaved case. This case appears on ghosted cells of a processor with rank
$p_j$ which only holds some of the cells to which the owning processor with
rank $p_i$ has assigned interleaved storage. This scenario can in general not
be avoided if only a single layer of ghost elements is present around the
locally owned subdomain of each MPI rank according to the algorithm
\cite{bbhk11}. As an example, consider the case of \texttt{SIMD\_WIDTH=2} for a cell at the corner with two additional ranks
according to Fig.~\ref{fig:ghost-mixed-stride}. Further, assume that the face $e_2\cap e_4$ is processed by processor $p_2$ and
face $e_2 \cap e_5$ by $p_3$, respectively. If the indices on $p_1$ are
interleaved between $e_1$ and $e_2$, this interleaved
access is also possible on $p_2$. However, $p_3$ has no knowledge of $e_1$, so
the ghost indices transformed to the MPI-local index space on $p_3$ appear to
have stride $1$ for $e_2$.
The \texttt{interleaved\_contiguous\_mixed\_strides} path is also used
for storing the case of partially filled SIMD lanes.

Data Structure \ref{alg:dof_info} proposes to store the indices of the degrees of
freedom as 32-bit unsigned integer numbers. We note that the index storage
refers to MPI-local numbering that starts with the locally owned range from
zero. Moreover, we use a global enumeration of degrees of freedom where
64-bit unsigned integers are possible for solving problems that have more than 4 billion
unknowns---but at no more than 4 billion unknowns per MPI process.

In Data Structure \ref{alg:dof_info}, we choose to redundantly store the start
indices of the cells on both sides of interior faces,
additionally to the start indices for cells. The indices on faces could
also be deduced from the stored cell indices and the face numbers
in \texttt{FaceInfo}. Since they only represent between 4 and 13 bytes for
each side of a face, it is usually more efficient to load this information
from a precomputed array than to recompute the face's index storage variant and start index during the operator
evaluation.
\begin{datastructure}[tb]
  \caption{\texttt{class DoFInfo}: Storage scheme for indices to degrees of freedom}
  \label{alg:dof_info}
  \begin{itemize}
  \setlength{\itemsep}{0pt}
  \setlength{\parsep}{0pt}
  \setlength{\parskip}{0pt}
  \item  \texttt{enum class IndexStorageVariants}:
  list of identifiers for different face storage types.
    \begin{itemize}
  \setlength{\itemsep}{0pt}
  \setlength{\parsep}{0pt}
  \setlength{\parskip}{0pt}
    \item \texttt{contiguous}: Contiguous degrees of freedom in each cell
      separately, storing only the start index. Read from
      \texttt{dof\_indices\_contiguous}.
    \item \texttt{interleaved\_contiguous}: Contiguous degree of freedom
      numbering within cells that is additionally interleaved over the SIMD
      arrays with unit stride, allowing for direct packed array access from
      the \texttt{SIMD\_WIDTH}-th index in \texttt{dof\_indices\_contiguous}.
    \item \texttt{interleaved\_contiguous\_strided}: Interleaved and
      contiguous storage within the cells. It has a fixed stride of
      \texttt{SIMD\_WIDTH} from one degree of freedom to the next, but a
      variable stride from one SIMD lane to the next.
    \item \texttt{interleaved\_contiguous\_mixed\_strides}: As opposed to
      \texttt{interleaved\_contiguous\_strided}, this storage scheme has a
      separate stride from one degree of freedom to next within the SIMD
      arrays. The strides are accessed from the vector
      \texttt{dof\_indices\_interleaved\_strides}.
    \item \texttt{full}: For handling continuous elements, generic storage
      with a separate index for each degree of freedom on each element is
      needed. Read from \texttt{dof\_indices}.
    \end{itemize}
  \item \texttt{std::vector<unsigned int> dof\_indices}: Generic index storage
    for generic elements, size equal to $n_{\text{cells}}k^d$, possibly
    including constraint information \cite{Kronbichler12}.
  \end{itemize}
  The following four fields keep three vectors that allow integrators on interior
  faces (indexed by 0), exterior faces (indexed by 1), and on cells (indexed by
  2) to directly access the appropriate location in the solution vectors
  without going over \texttt{FaceInfo}.
  \begin{itemize}
  \setlength{\itemsep}{0pt}
  \setlength{\parsep}{0pt}
  \setlength{\parskip}{0pt}
  \item \texttt{std::vector<IndexStorageVariants> index\_storage\_variant[3]}:
    Three vectors that define the detected index storage variant.
  \item \texttt{std::vector<unsigned int>
      dof\_indices\_contiguous[3]}: Three vectors storing the first index in a
    contiguous storage scheme from types
    \texttt{IndexStorageVariants::contiguous} through
    \texttt{IndexStorageVariants::interleaved\_contiguous\_mixed\_strides}.
  \item \texttt{std::vector<unsigned int>
      dof\_indices\_interleave\_strides[3]}: Three vectors storing the stride
    for each SIMD lane between consecutive degrees of freedom for
    \texttt{IndexStorageVariants::interleaved\_contiguous\_mixed\_strides}.
  \item \texttt{std::vector<unsigned char> n\_simd\_lanes\_filled[3]}: The
    number of SIMD lanes that are actually representing real data. For all
    lanes beyond the given number the vectorized integration kernels carry
    dummy data without read or write reference into vectors.
  \end{itemize}
\end{datastructure}

\subsection{Partitioning of face integrals with MPI parallelization}

The arithmetic work and memory access of face integrals is minimized if testing from both sides
of a face is done simultaneously, using the unique numerical flux for both sides and possible
terms from integration by parts. In this scheme, vector entries stored on neighboring
processors at the subdomain interfaces are needed. Assuming a balanced partitioning of
cells to be given from the domain decomposition, we divide the face integrals \emph{pairwise} on each interface
$p_i\cap p_j$ of two processors $p_i$ and $p_j$ to reach balanced work on faces. We set the following two
restrictions:
\begin{itemize}
\item If the set of shared faces $F_{ij}$ along the interface between $p_i$
  and $p_j$ contains two
  faces $f_a$ and $f_b$ referring to a single cell $e$ from processor $p_i$, we schedule both integrals on $p_j$. This approach
  reduces the amount of data sent, since the data from $e$ sent from $p_i$ to
  $p_j$ is used for two (or more) face integrals.
\item If a face $f\in F_{ij}$ is on an interface of different mesh levels
  (hanging node) where $p_i$ is on the coarser side and $p_j$ on the refined
  side, we schedule the integral on $p_j$. If $p_j$ holds all children, which
  is the most common case, we again reduce the data sent. Furthermore, this
  assumption allows to classify the off-processor side of faces as exterior
  cells $e^+$ according to \texttt{FaceInfo} in Algorithm~\ref{alg:face_info}
  and have subface interpolation only on the exterior side, allowing for a
  single \texttt{subface\_index} variable.
\end{itemize}
Since the number of cells is balanced and each processor gets half the faces
with each of its neighbors, the faces are also globally split evenly. This algorithm is cheaper than balancing faces more globally that involves more complicated strategies. We
observed experimentally that in all our setups, the number of locally computed
faces does not differ by more than up to a few dozens pf faces among all MPI ranks on
meshes more or less independent of the number of cells.

\subsection{Index access and ghosting}

For the integration kernel, remote
(ghosted) data on faces at processor boundaries must be imported before the respective face integrals are
issued. Furthermore, integral contributions computed on a remote processor
must finally be accumulated into the owner. Following the layout described in \cite{Kronbichler12}, we do this
by dedicated vector types that provide additional space beyond the locally
owned range to fit the ghost data. Note that we only communicate those cells where the face integrals demand data, not the full ghost layer. Outside the integration kernels, such as in time steppers or linear solvers, only the vector data of the locally owned range (one-to-one map over processors) is exposed and ghosted data is ignored.
Restricting the copy only to the ghosted entries is preferred because we aim for performance close
to the hardware limits,
where a copying of the whole vector together with the ghost entries would significantly increase computational costs.

\begin{figure}
  \centering
  \begin{tikzpicture}
    \begin{axis}[
      width=0.65\textwidth,
      height=0.47\textwidth,
      xlabel={Polynomial degree $p$},
      ylabel={Degrees of freedom per second},
      tick label style={font=\scriptsize},
      label style={font=\scriptsize},
      legend style={font=\scriptsize},
      legend pos = outer north east,
      title style={at={(1,0.963)},anchor=north east,draw=black,fill=white,font=\scriptsize},
      xmin=0,xmax=25,
      ymin=0,ymax=2.5e9,
      ytick={0,0.5e9,1e9,1.5e9,2e9,2.5e9},
      grid,
      cycle list name=colorGPL,
      mark size=1.8,
      semithick
      ]
      \addplot[gnuplot@darkblue,densely dashed,mark=o,every mark/.append style={solid}] table[x={degree}, y expr={\thisrowno{1}/\thisrowno{3}}] {\matvecBdwVeryLarge};
      \addlegendentry{advection slim very large};
      \addplot[gnuplot@darkblue,mark=o] table[x={degree}, y expr={\thisrowno{1}/\thisrowno{3}}] {\matvecBasesAdveBdw};
      \addlegendentry{advection slim};
      \addplot[gnuplot@darkblue,mark=o,densely dotted,every mark/.append style={solid}] table[x={degree}, y expr={\thisrowno{1}/\thisrowno{5}}] {\matvecBasesAdveBdw};
      \addlegendentry{advection full};
      \addplot[red!80!white,densely dashed,mark=x,every mark/.append style={solid}] table[x={degree}, y expr={\thisrowno{1}/\thisrowno{2}}] {\matvecBdwVeryLarge};
      \addlegendentry{Laplacian slim very large};
      \addplot[red!80!white,mark=x] table[x={degree}, y expr={\thisrowno{1}/\thisrowno{3}}] {\matvecBasesLaplBdw};
      \addlegendentry{Laplacian slim};
      \addplot[red!80!white,mark=x,densely dotted,every mark/.append style={solid}] table[x={degree}, y expr={\thisrowno{1}/\thisrowno{7}}] {\matvecBasesLaplBdw};
      \addlegendentry{Laplacian full};
    \end{axis}
  \end{tikzpicture}
  \caption{Influence of the MPI communication pattern on throughput as a function of
    the polynomial degree on 28 Broadwell cores for the Laplacian on a Hermite-like basis. The
    problem sizes are between 8 and 57 million DoFs and jump to
    the next smaller grid at $p=3$, $p=6$, $p=12$, and $p=24$, increasing the
    proportion of cells at processor boundaries from 4\% at $p\leq 2$ to more
    than 80\% at $p\geq 24$. The dashed and dashdotted lines display the
    throughput at sizes above 160 million degrees of freedom with the slim
    data communication.}
\label{fig:throughput_mpi}
\end{figure}
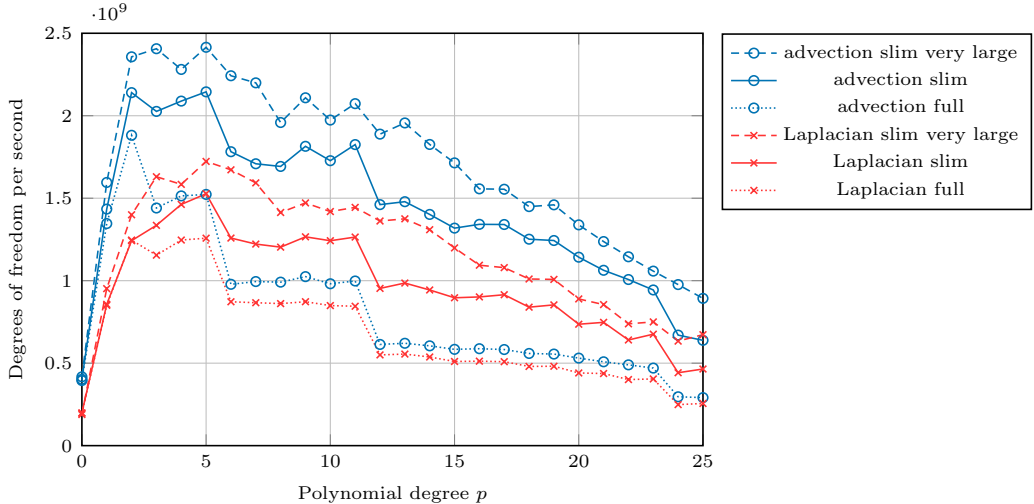

Fig.~\ref{fig:throughput_mpi} lists the throughput of the full matrix-vector
product including the MPI communication. The communication of all the degrees
of freedom of remote cells needed for face integrals is labeled as ``full'' in the figure. To ensure
comparable problem sizes, we use Cartesian meshes whose size varies with the
polynomial degree, namely a $128^3$ mesh for degrees $p=1,2$, a $64^3$ mesh
for $3\leq p\leq 5$, a $32^3$ mesh for $6\leq p \leq 11$, a $16^3$ mesh for
$12\leq p\leq 23$, and an $8^3$ mesh for $p\geq 24$. For this choice, the
problem has between 8 and 57 million degrees of
freedom. Fig.~\ref{fig:throughput_mpi} reveals sudden drops in performance
when the mesh size is reduced. At these points, the volume-to-surface ratio
for the next smaller mesh level gets lower and a larger proportion of cells
must be exchanged. Note that MPI communication in this experiment is in fact
within a shared-memory architecture, so the cost is related to some
\texttt{memcpy} kernels as well as pack and unpack routines which are memory
bandwidth bound. This part consumes more than 40\% of the operator evaluation
time for $6\leq p\leq 11$, and we have verified that it indeed runs at full
memory speed of $110$ GB/s on Broadwell.

The algorithm of sending all data of a cell can be improved by observing that the
\texttt{face\_normal\-\_interpolation} kernel will only touch some of the
degrees of freedom of a cell for certain bases and derivatives, namely those where $S_\mathrm{f}$ and possibly $D_\mathrm{f}$ of Eq.~\eqref{eq:face_integral} are non-zero. In this case,
only the relevant part of the cell's vector entries must be communicated
and packed/unpacked. Algorithm~\ref{alg:mpi_slim} presents the
algorithmic setup. The algorithm keeps the integration and vector access
routines agnostic of this fact: we provide storage for all the degrees of freedom of a cell but only populate some of the entries with data. Copying the slim indices adds another indirection to the ``pack'' stage
of the MPI data exchange because the data to be sent is not contiguous in memory.
Fortunately, the code pattern is exactly the
same as the selection of some entries among the locally owned part
for data transfer, so existing implementations can be readily extended. Note
that Algorithm \ref{alg:mpi_slim} cannot be directly implemented via
general-purpose vector classes with MPI facilities such as PETSc
\cite{petsc-user-ref} or Trilinos' Epetra and Tpetra
\cite{trilinos,trilinos-web-page} without deep vector copies, which is why we
use our own specialized implementation inside \texttt{deal.II} that is
interlinked with the needs of the integrators.

\begin{algorithm}[tb]
  \caption{Slim MPI communication}
  \label{alg:mpi_slim}
\begin{itemize}
  \setlength{\itemsep}{0pt}
  \setlength{\parsep}{0pt}
  \setlength{\parskip}{0pt}
\item Required input:
\begin{itemize}
  \setlength{\itemsep}{0pt}
  \setlength{\parsep}{0pt}
  \setlength{\parskip}{0pt}
\item Precomputed type of basis: nodal at boundary, Hermite-type, generic.
\item Which terms are required for inner face integrals (only cell terms, values on face, values and
  first derivatives on face)?
\end{itemize}
\item \texttt{update\_ghost\_values} fills the ghosted data, \texttt{compress}
  accumulates integral contributions to the respective entries at the
  owner. The communication is established according to the following three
  options:
\begin{itemize}
  \setlength{\itemsep}{0pt}
  \setlength{\parsep}{0pt}
  \setlength{\parskip}{0pt}
\item If no face integrals, do nothing in DG (or cell-only ghost exchange
  of continuous elements).
\item If only face values and basis nodal at boundary, send only one single
  layer of $k^{d-1}$ degrees of freedom per cell at the interface.
\item If only up to first derivatives on face and Hermite-type basis, send
  only the two layers representing values and normal derivatives with
  $2k^{d-1}$ degrees of freedom per cell at the interface.
\end{itemize}
We provide storage for all degrees of freedom of the ghosted cell in the
vector but only fill the entries where the interpolation matrices $S_\text{f}$
and $D_\text{f}$ are nonzero (avoids separate code paths).
\end{itemize}
\end{algorithm}

The proposed slim MPI communication leads to simpler code than an
alternative concept that uses a face-separate storage scheme at
processor boundaries that runs \texttt{face\_normal\_interpolation} before communication.  A different storage scheme for data from ghosted cells
with other strides would lead to a lane divergence in case not all
faces in the SIMD array are adjacent to a ghosted cell: the lanes
without ghosted cells would still need to perform an interpolation step whereas the ones
representing ghosted cells would not.

Fig.~\ref{fig:throughput_mpi} shows that the slim MPI communication of
Algorithm~\ref{alg:mpi_slim} improves throughput by more than 30\% for the
Laplacian with $p=6$ at 8.9 million degrees of freedom and by almost 70\% for
the advection operator with $p=6$. The drops in
throughput when going to smaller meshes with larger surfaces at processor
boundaries are still visible for the slim implementation but much less
pronounced. As a point of reference, Fig.~\ref{fig:throughput_mpi} also
contains data from experiments on problems on a mesh that is more than 20
times larger where the slim MPI communication amounts to less than 10\% of the
operator evaluation time.

In Table~\ref{tab:face-separate} we compare the throughput of the proposed
slim MPI data exchange for the 3D Laplacian on a Hermite-type basis with a
face-separate storage as used e.g.~in \cite{Hindenlang12,Kronbichler17}. The
face-separate storage breaks the face integrals into a separate loop and
handles the cell--face exchange through a global data structure for faces. This
setup simplifies the MPI data exchange and allows for easier threading, but at
a much higher memory access since the result vector must be accessed in
two separate loops and the face storage in three loops
\cite{Kronbichler17}.
For face-separate storage a basis with collocation of nodal points and quadrature points is more advantageous because it can skip the \texttt{basis\_change} algorithm in cell integrals.
The results highlight that the proposed method with a
single loop for cell and face integrals is almost twice as fast for degree
$p=5$ and still 25\% faster for $p=11$.
The face-separate storage only becomes
superior at $p > 15$ where the advantages of collocated node and quadrature points
get significant. Note that the face-separate storage scheme
runs at full memory throughput with $>90\, \text{GB/s}$ for all degrees up to
$p=11$, whereas the proposed scheme does not utilize the full RAM bandwidth,
despite being substantially faster. Thus, even an idealized version of the face-separate storage would be necessarily slower than our proposed implementation.

\begin{table}
  \caption{Throughput and measured memory throughput for evaluation of the 3D
    Laplacian on $2\times 14$ Broadwell cores with the proposed implementation
    on a Hermite-like basis against scheme with separate face storage on a
    with collocated nodal and quadrature points that minimizes arithmetic operations.}
  \label{tab:face-separate}
  \smallskip
{
    \small\strut\hfill
  \begin{tabular}{lccccccc}

    \hline
    polynomial degree $p$ & 2 & 3 & 5 & 8 & 11 & 15 & 20\\
    \hline
                          & \multicolumn{7}{c}{million degrees of freedom per second, MDoF/s}\\
    proposed scheme & 1243 & 1336 & 1527 & 1203 & 1264 & 896 & 736 \\
    face-separate storage & 551 & 646 & 897 & 1014 & 1077 & 901 & 912 \\
    \hline
                          & \multicolumn{7}{c}{measured memory throughput, GB/s}\\
    proposed scheme & 62.2 & 68.6 & 74.0 & 71.1 & 64.4 & 61.1 & 63.6 \\
    face-separate storage & 106 & 106 & 107 & 102 & 90.7 & 75.1 & 72.6 \\
    \hline
  \end{tabular}
  \hfill\strut
}
\end{table}

\begin{figure}
  \centering
  \begin{tikzpicture}
    \begin{axis}[
      width=0.5\textwidth,
      height=0.5\textwidth,
      xlabel={Polynomial degree $p$},
      ylabel={Time per billion degrees of freedom [s]},
      tick label style={font=\scriptsize},
      label style={font=\scriptsize},
      legend to name=legendIntegralDetails,
      legend columns=3,
      legend style={font=\scriptsize},
      title style={at={(1,0.963)},anchor=north east,draw=black,fill=white,font=\scriptsize},
      title={Problem size 1M to 7.1M},
      xmin=1,xmax=25,
      ymin=0,ymax=2.4,
      xtick={1,5,9,13,17,21,25},
      grid,
      cycle list name=colorGPL,
      mark size=1.8,
      stack plots=y,
      area style,
      semithick
      ]
      \addplot table[x={degree}, y expr={\thisrowno{8}/\thisrowno{1}*1e9}] {\matvecDetailsSma}
      \closedcycle;
      \addlegendentry{Bookkeeping and control logic};
      \addplot table[x={degree}, y expr={\thisrowno{6}/\thisrowno{1}*1e9}] {\matvecDetailsSma}
      \closedcycle;
      \addlegendentry{Sum factorization cells};
      \addplot table[x={degree}, y expr={\thisrowno{7}/\thisrowno{1}*1e9}] {\matvecDetailsSma}
      \closedcycle;
      \addlegendentry{q-point operation cells};
      \addplot table[x={degree}, y expr={\thisrowno{2}/\thisrowno{1}*1e9}] {\matvecDetailsSma}
      \closedcycle;
      \addlegendentry{Vector access face integrals};
      \addplot table[x={degree}, y expr={\thisrowno{4}/\thisrowno{1}*1e9}] {\matvecDetailsSma}
      \closedcycle;
      \addlegendentry{Sum factorization faces};
      \addplot table[x={degree}, y expr={\thisrowno{5}/\thisrowno{1}*1e9}] {\matvecDetailsSma}
      \closedcycle;
      \addlegendentry{q-point operation faces};
      \addplot[draw=red!60!blue!70!white,fill=red!60!blue!20!white] table[x={degree}, y expr={(\thisrowno{9}-\thisrowno{2}-\thisrowno{4}-\thisrowno{5}-\thisrowno{6}-\thisrowno{7}-\thisrowno{8})/\thisrowno{1}*1e9}] {\matvecDetailsSma}
      \closedcycle;
      \addlegendentry{Local operation load imbalance};
      \addplot[draw=blue!70!green!60!white,fill=blue!70!green!20!white] table[x={degree}, y expr={(\thisrowno{10}-\thisrowno{9})/\thisrowno{1}*1e9}] {\matvecDetailsSma}
      \closedcycle;
      \addlegendentry{MPI comm.: (un)pack + \texttt{memcpy}};
      \addplot[line legend,sharp plot,black,mark=diamond,stack plots=false] table[x={degree}, y expr={(\thisrowno{11})/\thisrowno{1}*1e9}] {\matvecDetailsSma};
      \addlegendentry{Compute without timers};
    \end{axis}
  \end{tikzpicture}
  ~
  \begin{tikzpicture}
    \begin{axis}[
      width=0.5\textwidth,
      height=0.5\textwidth,
      xlabel={Polynomial degree $k$},
      ylabel={Time per billion degrees of freedom [s]},
      tick label style={font=\scriptsize},
      label style={font=\scriptsize},
      title style={at={(1,0.963)},anchor=north east,draw=black,fill=white,font=\scriptsize},
      title={Problem size 160M to 1.1B},
      xmin=1,xmax=25,
      xtick={1,5,9,13,17,21,25},
      ymin=0,ymax=2.4,
      grid,
      cycle list name=colorGPL,
      mark size=1.8,
      stack plots=y,
      area style,
      semithick
      ]
      \addplot table[x={degree}, y expr={\thisrowno{8}/\thisrowno{1}*1e9}] {\matvecDetailsLar}
      \closedcycle;
      \addplot table[x={degree}, y expr={\thisrowno{6}/\thisrowno{1}*1e9}] {\matvecDetailsLar}
      \closedcycle;
      \addplot table[x={degree}, y expr={\thisrowno{7}/\thisrowno{1}*1e9}] {\matvecDetailsLar}
      \closedcycle;
      \addplot table[x={degree}, y expr={\thisrowno{2}/\thisrowno{1}*1e9}] {\matvecDetailsLar}
      \closedcycle;
      \addplot table[x={degree}, y expr={\thisrowno{4}/\thisrowno{1}*1e9}] {\matvecDetailsLar}
      \closedcycle;
      \addplot table[x={degree}, y expr={\thisrowno{5}/\thisrowno{1}*1e9}] {\matvecDetailsLar}
      \closedcycle;
      \addplot[draw=red!60!blue!70!white,fill=red!60!blue!20!white] table[x={degree}, y expr={(\thisrowno{9}-\thisrowno{2}-\thisrowno{4}-\thisrowno{5}-\thisrowno{6}-\thisrowno{7}-\thisrowno{8})/\thisrowno{1}*1e9}] {\matvecDetailsLar}
      \closedcycle;
      \addplot[draw=blue!70!green!60!white,fill=blue!70!green!20!white] table[x={degree}, y expr={(\thisrowno{10}-\thisrowno{9})/\thisrowno{1}*1e9}] {\matvecDetailsLar}
      \closedcycle;
      \addplot[line legend,sharp plot,black,mark=diamond,stack plots=false] table[x={degree}, y expr={(\thisrowno{11})/\thisrowno{1}*1e9}] {\matvecDetailsLar};
    \end{axis}
  \end{tikzpicture}
  \\
  \ref{legendIntegralDetails}
  \caption{Breakdown of computation times on different phases for the 3D Laplacian
    for a small sized experiment at 1--7.1 million DoFs and a large experiment
    at 160 million to 1.1 billion DoFs. The timings have been recorded on $2\times 14$ Broadwell cores and rescaled to the time for a fixed
    fictive size of 1 billion ($10^9$) degrees of freedom.}
\label{fig:timings_phases}
\end{figure}
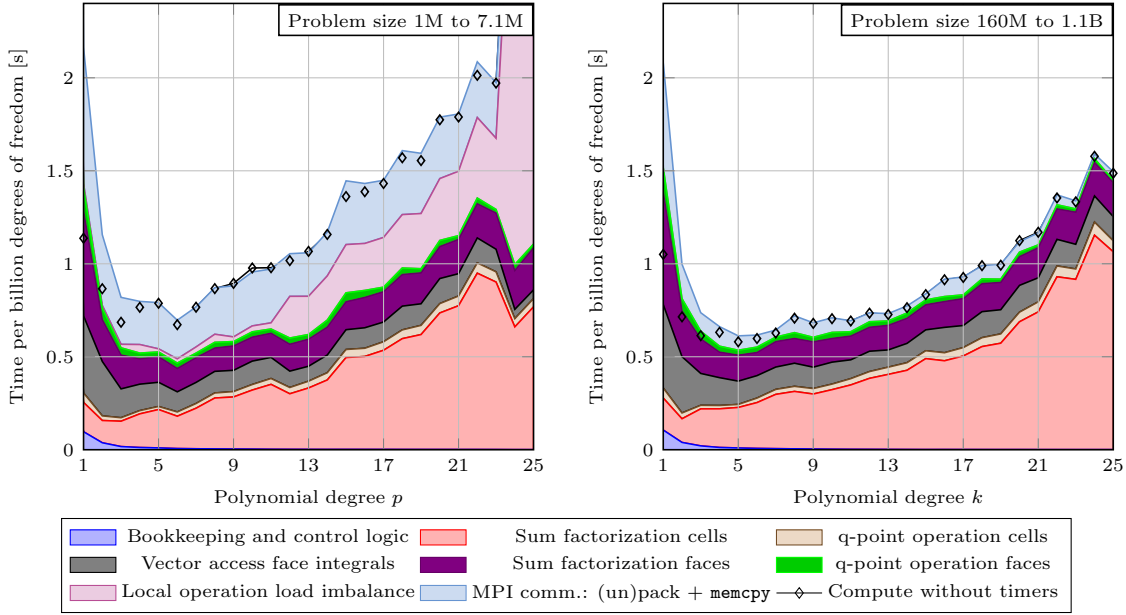

Fig.~\ref{fig:timings_phases} specifies the cost of the various algorithmic
components of the DG operator evaluation. Timings are based on the RDTSC timer
register of Intel processors that have small enough overhead to be placed in
inner loops. When compared to the compute-only part from
Fig.~\ref{fig:dg_ops_arch}, we observe a drop in performance by approximately
a factor of 2 for the Laplacian for $2\leq p\leq 10$. Clearly, the MPI
communication (including \texttt{memcpy} and pack/unpack) takes a significant
share of time for the smaller computation. Sum factorization on cells which is
of complexity $\mathcal O(p)$ in the number of degrees of freedom gets more
expensive for higher degrees, but it takes less than 50\% of the evaluation
time for all degrees $p\leq 10$. To show uncertainties and influence
of timers on out-of-order execution that are significant at $p<4$, we also
include reference runs with disabled timers, displayed by diamonds.

\begin{Remark}
  Note that the above experiments are performed in shared memory, so
  the MPI data exchange boils down to some form of \texttt{memcpy} operations
  in the MPI implementation besides the pack/unpack parts. Thus, there is no direct advantage of overlapping
  the communications with computations and the transfer cost does show up in our
  profiles, as opposed to communication over an Infiniband-like fabric that
  indeed is overlapped with our implementation.
\end{Remark}

\subsection{Performance on Haswell, Broadwell and Knights Landing}

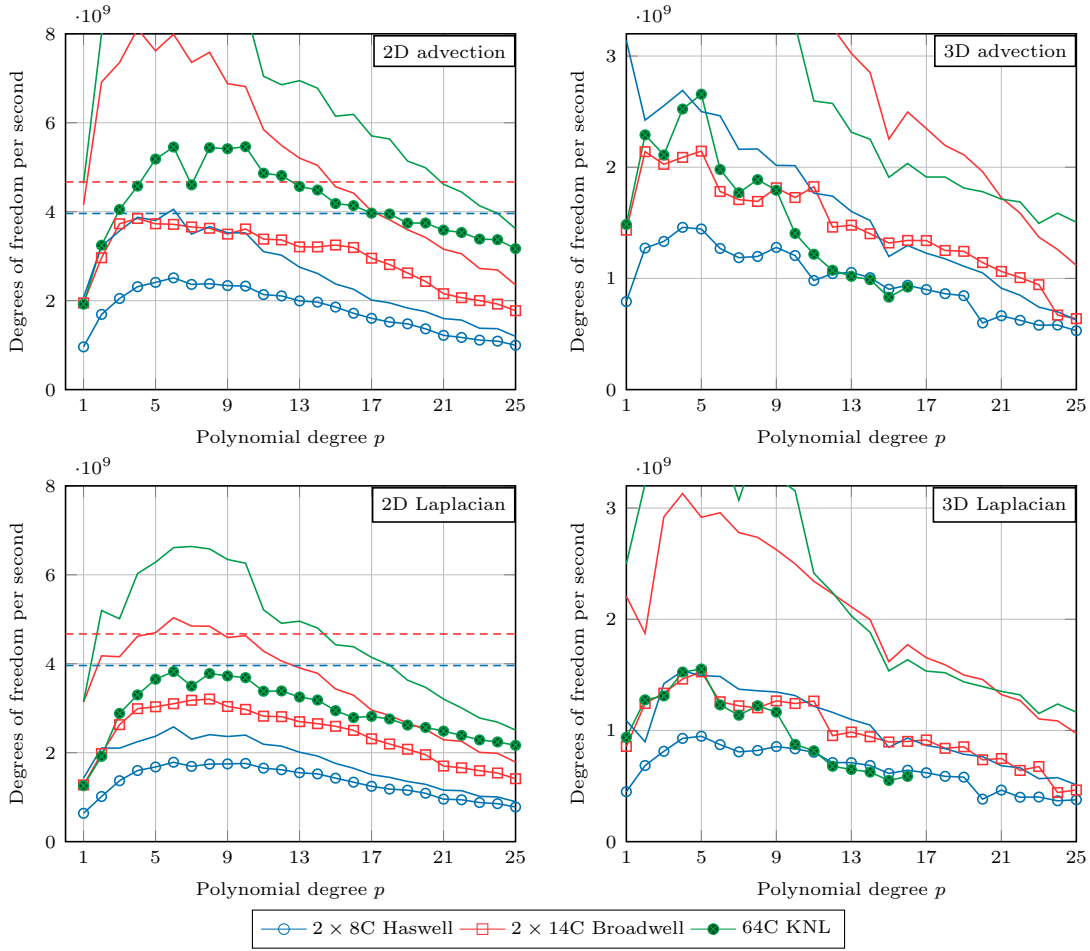
\begin{figure}
  \centering
  \begin{tikzpicture}
    \begin{axis}[
      width=0.5\textwidth,
      height=0.42\textwidth,
      xlabel={Polynomial degree $p$},
      ylabel={Degrees of freedom per second},
      tick label style={font=\scriptsize},
      label style={font=\scriptsize},
      legend to name=legendMatvecSystems,
      legend columns=3,
      legend style={font=\scriptsize},
      title style={at={(1,0.952)},anchor=north east,draw=black,fill=white,font=\scriptsize},
      title={2D advection},
      xmin=0,xmax=25,
      xtick={1,5,9,13,17,21,25},
      ymin=0,ymax=8e9,
      grid,
      cycle list name=colorGPL,
      mark size=1.8,
      semithick
      ]
      \addplot table[x={degree}, y expr={\thisrowno{1}/\thisrowno{3}}] {\matvecSystemsTwoD};
      \addlegendentry{$2\times 8$C Haswell};
      \addplot table[x={degree}, y expr={\thisrowno{1}/\thisrowno{6}}] {\matvecSystemsTwoD};
      \addlegendentry{$2\times 14$C Broadwell};
      \addplot table[x={degree}, y expr={4*\thisrowno{1}/\thisrowno{9}}] {\matvecSystemsTwoD};
      \addlegendentry{64C KNL};
      \addplot[gnuplot@darkblue,solid] table[x={deg}, y={HswConvect}] {\dgDofsTwo};
      \addplot[red!80!white,solid] table[x={deg}, y={BdwConvect}] {\dgDofsTwo};
      \addplot[gnuplot@green,solid] table[x={deg}, y={KnlConvect}] {\dgDofsTwo};
      \addplot[gnuplot@darkblue,densely dashed] coordinates {
        (0,3.96e+9) (25,3.96e+9)
      };
      \addplot[red!80!white,densely dashed] coordinates {
        (0,4.67e+9) (25,4.67e+9)
      };
    \end{axis}
  \end{tikzpicture}
  ~
  \begin{tikzpicture}
    \begin{axis}[
      width=0.5\textwidth,
      height=0.42\textwidth,
      xlabel={Polynomial degree $p$},
      ylabel={Degrees of freedom per second},
      tick label style={font=\scriptsize},
      label style={font=\scriptsize},
      legend style={font=\scriptsize},
      title style={at={(1,0.952)},anchor=north east,draw=black,fill=white,font=\scriptsize},
      title={3D advection},
      xmin=1,xmax=25,
      xtick={1,5,9,13,17,21,25},
      ymin=0,ymax=3.2e9,
      grid,
      cycle list name=colorGPL,
      mark size=1.8,
      semithick
      ]
      \addplot table[x={degree}, y expr={\thisrowno{1}/\thisrowno{3}}] {\matvecSystems};
      \addplot table[x={degree}, y expr={\thisrowno{1}/\thisrowno{6}}] {\matvecSystems};
      \addplot table[x={degree}, y expr={\thisrowno{1}/\thisrowno{9}}] {\matvecSystems};
      \addplot[gnuplot@darkblue,solid] table[x={deg}, y={HswConvect}] {\dgDofsThree};
      \addplot[red!80!white,solid] table[x={deg}, y={BdwConvect}] {\dgDofsThree};
      \addplot[gnuplot@green,solid] table[x={deg}, y={KnlConvect}] {\dgDofsThree};
      \addplot[gnuplot@darkblue,densely dashed] coordinates {
        (0,3.96e+9) (25,3.96e+9)
      };
      \addplot[red!80!white,densely dashed] coordinates {
        (0,4.67e+9) (25,4.67e+9)
      };
    \end{axis}
  \end{tikzpicture}
  \\
  \begin{tikzpicture}
    \begin{axis}[
      width=0.5\textwidth,
      height=0.42\textwidth,
      xlabel={Polynomial degree $p$},
      ylabel={Degrees of freedom per second},
      tick label style={font=\scriptsize},
      label style={font=\scriptsize},
      title style={at={(1,0.952)},anchor=north east,draw=black,fill=white,font=\scriptsize},
      title={2D Laplacian},
      xmin=0,xmax=25,
      xtick={1,5,9,13,17,21,25},
      ymin=0,ymax=8e9,
      grid,
      cycle list name=colorGPL,
      mark size=1.8,
      semithick
      ]
      \addplot table[x={degree}, y expr={\thisrowno{1}/\thisrowno{2}}] {\matvecSystemsTwoD};
      \addplot table[x={degree}, y expr={\thisrowno{1}/\thisrowno{5}}] {\matvecSystemsTwoD};
      \addplot table[x={degree}, y expr={4*\thisrowno{1}/\thisrowno{8}}] {\matvecSystemsTwoD};
      \addplot[gnuplot@darkblue,solid] table[x={deg}, y={HswLaplace}] {\dgDofsTwo};
      \addplot[red!80!white,solid] table[x={deg}, y={BdwLaplace}] {\dgDofsTwo};
      \addplot[gnuplot@green,solid] table[x={deg}, y={KnlLaplace}] {\dgDofsTwo};
      \addplot[gnuplot@darkblue,densely dashed] coordinates {
        (0,3.96e+9) (25,3.96e+9)
      };
      \addplot[red!80!white,densely dashed] coordinates {
        (0,4.67e+9) (25,4.67e+9)
      };
    \end{axis}
  \end{tikzpicture}
  ~
  \begin{tikzpicture}
    \begin{axis}[
      width=0.5\textwidth,
      height=0.42\textwidth,
      xlabel={Polynomial degree $p$},
      ylabel={Degrees of freedom per second},
      tick label style={font=\scriptsize},
      label style={font=\scriptsize},
      legend style={font=\scriptsize},
      title style={at={(1,0.952)},anchor=north east,draw=black,fill=white,font=\scriptsize},
      title={3D Laplacian},
      xmin=1,xmax=25,
      xtick={1,5,9,13,17,21,25},
      ymin=0,ymax=3.2e9,
      grid,
      cycle list name=colorGPL,
      mark size=1.8,
      semithick
      ]
      \addplot table[x={degree}, y expr={\thisrowno{1}/\thisrowno{2}}] {\matvecSystems};
      \addplot table[x={degree}, y expr={\thisrowno{1}/\thisrowno{5}}] {\matvecSystems};
      \addplot table[x={degree}, y expr={\thisrowno{1}/\thisrowno{8}}] {\matvecSystems};
      \addplot[gnuplot@darkblue,solid] table[x={deg}, y={HswLaplace}] {\dgDofsThree};
      \addplot[red!80!white,solid] table[x={deg}, y={BdwLaplace}] {\dgDofsThree};
      \addplot[gnuplot@green,solid] table[x={deg}, y={KnlLaplace}] {\dgDofsThree};
      \addplot[gnuplot@darkblue,densely dashed] coordinates {
        (0,3.96e+9) (25,3.96e+9)
      };
      \addplot[red!80!white,densely dashed] coordinates {
        (0,4.67e+9) (25,4.67e+9)
      };
    \end{axis}
  \end{tikzpicture}
  \\
  \ref{legendMatvecSystems}
  \caption{Comparison of the throughput on the 2D and 3D advection operator
    and Laplacian on Haswell,
    Broadwell and KNL. The measurements are compared to the theoretical
    throughput of the kernels from Fig.~\ref{fig:dg_ops_arch} (solid lines of respective color)
    and the memory bandwidth of 95 GB/s on Haswell, 112 GB/s on
    Broadwell (dashed lines) and 450 GB/s on KNL (outside shown range) for the idealized setting of two vector reads
    and one vector write.}
\label{fig:timings_dg}
\end{figure}

Fig.~\ref{fig:timings_dg} shows our main result, the
performance of the presented implementation on the DG
discretization of the advection equation \eqref{eq:advection} and the
DG discretization of the Laplacian
\eqref{eq:poisson_sip}, including the full code with MPI data exchange. We
compare the actually achieved performance (lines with marks) to the
theoretical arithemtic performance limit established in
Fig.~\ref{fig:dg_ops_arch} and the memory bandwidth limit. The actual
throughput is between $0.41\times$ and $0.77\times$ the arithmetic throughput
numbers for the 2D advection operator at $2\leq p\leq 15$ for all three
architectures and mostly around $0.55\times$. For the 2D Laplacian, typical
values are at $0.69\times$ of the theoretical throughput on Haswell,
$0.63\times$ on Broadwell, and $0.55\times$ on KNL. In analogy to
Fig.~\ref{fig:timings_phases}, the gap in performance can be accounted to the
MPI communication (around 5--8\% of operator evaluation in 2D), the gather
access pattern at half the faces (around 15\% of operator evaluation
time)\footnote{The gather operation occurs on slightly more than half of the
  faces: in the computations shown in Fig.~\ref{fig:timings_phases} around
  $45\ldots 50\%$ of the faces involve direct array access according to
  \texttt{interleaved\_contiguous} type, $40\ldots 50\%$ involve the
  \texttt{interleaved\_contiguous\_strided} variant, and the remainder $10\%$
  have mixed strides.} and, for degrees less than 3, to overhead in accessing
the storage data structures such as Data Structure~\ref{alg:dof_info}. If we take into account these factors,
performance is within 10\% of the possible roofline on Haswell and Broadwell
and within 20\% on KNL. In the likwid analysis, the remaining performance
penalties were mostly identified to be memory stalls due to wait for L2/L3
caches or memory from imperfect prefetching. To account for these effects, a
more detailed look at the memory hierarchy would be necessary beyond the
roofline concepts applied here, such as the execution cache memory performance
model \cite{Hager11}.

In three space dimensions, we observe a considerably larger gap
between the model and the actual performance. On Haswell, we record a
throughput of $0.56\times$ and $0.63\times$ the value for the compute-only
case for $2\leq k\leq 10$ for advection and the Laplacian, respectively, the
numbers of Broadwell are $0.41\times$ and $0.49\times$, and for KNL we
recorded $0.39\times$ in both cases. As documented by
Fig.~\ref{fig:timings_phases}, up to a third of the compute time is consumed
in the MPI communication and pack/unpack already for $p$ between 6 and 11. If
subtracting the data exchange, the Laplacian reaches around 1.8 to 2.1 billion
degrees of freedom per second on 28 Broadwell cores and 64 KNL cores for
$4\leq k\leq 9$, i.e., more than 70\% of the compute performance of
Fig.~\ref{fig:timings_phases}. Besides the cost for gather and scatter
operations, the performance gap not further quantified and attributable to
memory stalls is around 20\% in 3D.

The algorithms are equally beneficial for large-scale
parallel simulations as shown in Table~\ref{tab:scaling}. We record almost
ideal strong scaling until around 0.7\,{ms} and reasonable scaling
down to 0.3\,{ms}. The main reason for the loss in efficiency above
0.7\,{ms} is the larger proportion of pack/unpack and \texttt{memcpy}
routines in the communication reported in
Fig.~\ref{fig:timings_phases}. The proposed algorithms have delivered excellent performance to even larger scale up to almost 10,000
nodes (147,456 cores) in \cite{Kronbichler16b,Krank17}.

\begin{table}
  \caption{Strong scaling of experiment on 3D Cartesian mesh with 262,144 cells and $p=5$ (56.6 million degrees of freedom) on Xeon E5-2697 v3 ($2\times 14$ cores at 2.6 GHz) based cluster on up to 512 nodes with 14,336 cores. Numbers are reported as the absolute run time in milliseconds $[\mathrm{ms}]$ of a matrix-vector product including communication and in terms of throughput measured as million degrees of freedom per second and core (MDoF/s/core) reporting the parallel efficiency.}
  \label{tab:scaling}
  \smallskip
{
    \small\strut\hfill
\begin{tabular}{lcccccccccc}
  \hline
  \# nodes & 1 & 2 & 4 & 8 & 16 & 32 & 64 & 128 & 256 & 512\\
  \hline
  & \multicolumn{9}{c}{double precision}\\
  time $[\mathrm{ms}]$ & 41.6 & 21.4 & 11.7 & 6.23 & 3.12 & 1.43 & 0.798 & 0.495 & 0.409 & 0.236\\
  MDoF/s/core & 48.7 & 47.2 & 43.3 & 40.6 & 41.6 & 44.1 & 39.6 & 31.9 & 19.3 & 16.8 \\
  \hline
  & \multicolumn{9}{c}{single precision}\\
  time $[\mathrm{ms}]$ & 21.9 & 11.8 & 6.12 & 3.07 & 1.48 & 0.814 & 0.494 & 0.324 & 0.285 & 0.177\\
  MDoF/s/core & 92.4 & 86.0 & 82.6 & 82.3 & 85.1 & 77.6 & 64.0 & 48.7 & 27.7 & 23.3\\
  \hline
\end{tabular}
\hfill\strut
}
\end{table}

\section{Representation of geometry}\label{sec:geometry}

In Algorithm~\ref{alg:prototype_dof_op}, we assumed the
inverse Jacobian $\mathcal J^{(e)}$ of the transformation from unit to real
cell to be given. The computation of integrals also involves the
determinant of the Jacobian and normal vectors $\ve n$ derived from the
Jacobian. The Jacobian is often defined as the derivative of a piecewise $m$-th degree
polynomial description of the geometry through some
points which we refer to as mapping support points $\ve x_{\text{msp}}^{(i)}$,
$i=1,\ldots,n_\text{points}$, but it can also be defined by analytical tangent
vectors on the geometry \cite{dealII85}. Extending over a short discussion in
\cite{Kronbichler12}, a high-performance implementation for arbitrary
geometries can select between at least four main variants:
\begin{itemize}
\item[{(G1)}] Storage of mapping support point locations $\ve x_{\text{msp}}^{(i)}$
  for all indices $i$ in the mesh with usual indirect addressing of continuous finite elements
  and subsequent tensor product evaluation. This involves transfer of
  \textit{somewhat less than 3 double precision values} per quadrature point for
  isoparametric mappings, depending on the polynomial degree which determines
  the weight on the higher valence entities, i.e., mesh vertices, edges, and
  faces.
\item[{(G2)}] Storage of all quadrature points $\ve x_{\text{qp}}^{(q)}$ in
  physical space, from which the Jacobian can be computed by a
  collocation derivative. This needs \textit{3 double precision values} per quadrature
  point in 3D. For face integrals, separate data is needed except for
  Gauss--Lobatto like integration rules.
\item[{(G3)}] Pre-computation of $\mathcal J^{(e)}$ in all quadrature points
  of the mesh; for face integrals,
  separate data is needed. Storing the inverse Jacobian and the determinant of
  the Jacobian involves transfer of \textit{10 double precision values} per quadrature point in 3D.
\item[{(G4)}] Pre-computation of the effective coefficient in the particular equation
  at hand, e.g. a symmetric $d\times d$ tensor
  $ \left(\mathcal
    J^{(e)}\right)^{-1}\left(\mathcal J^{(e)}\right)^{-\mathrm T}\text{det}(\mathcal J^{(e)}) $ for the
  cell term of the Laplacian (\textit{6 double precision values} in 3D) or the vector
  $ \left(\mathcal J^{(e)}\right)^{-1}\ve c(\ve x)\text{det}(\mathcal J^{(e)})
  $ for the advection equation (\textit{3 double precision values} in 3D), a technique used
  e.g.~in Nek5000 \cite{nek5000-web-page}.
\end{itemize}
Furthermore, coefficients such as
$\ve c\left(\hat{\ve x}^{(e)}(\boldsymbol \xi)\right)$ in Algorithm
\eqref{alg:prototype_dof_op} can be pre-computed and loaded during
the operator evaluation or computed on the fly based on the location of the
quadrature point $\ve x_{\text{qp}}^{(q)}$. The pre-computed variants can also
be combined with simple memory compressions, such as the constant-Jacobian
case on affine meshes \cite{Kronbichler12} or constant-in-one-direction case on extruded
meshes. The results in the previous section have shown that the operator evaluation is
usually compute bound on Cartesian meshes, in particular on the Haswell and KNL systems.
Since the first two options (G1) and (G2) involve additional computations,
it is not clear a priori whether to prefer tabulation with its higher memory transfer
or rather computation on the fly.

For a generic software package, a further factor that must be taken into
account are possible computations or memory access in user code. Our
experience is that these tend to rather introduce compute-bound patterns
including table lookups, branches, or computations, such that we prefer the
pre-computed variants. In \texttt{deal.II}, we use a sparse-matrix-like storage
scheme that allows different cells and faces to hold arrays of different
lengths, for example to flexibly switch between Cartesian meshes or fully
curved meshes or hp-adaptive data structures.  We propose to keep different
arrays for quantities such as normal vectors, Jacobians, and Jacobian
determinants, so that only the data that is
actually needed for a particular operator evaluation is loaded from memory. We observed that attempts to interleave storage
of different quantities closer together has a negative impact on performance
because data may overlap in cache lines or hardware prefetchers would eagerly
load unnecessary data.

The variants (G3) and (G4) result in different code layouts, respectively: (G3) allows for
the definition of arbitrary weak forms and integration of nonlinear terms in
general scenarios by separate control over the operators on trial functions
and test functions (e.g.~\texttt{get\_value}, \texttt{get\_gradient}, or
\texttt{test\_by\_gradient}). The variant (G4) hardcodes the differential
operator in the coefficient. Certain operators like the Laplacian or advection
are treated more efficiently with the latter approach. To satisfy these
opposing demands, we allow for both in our generic implementation.

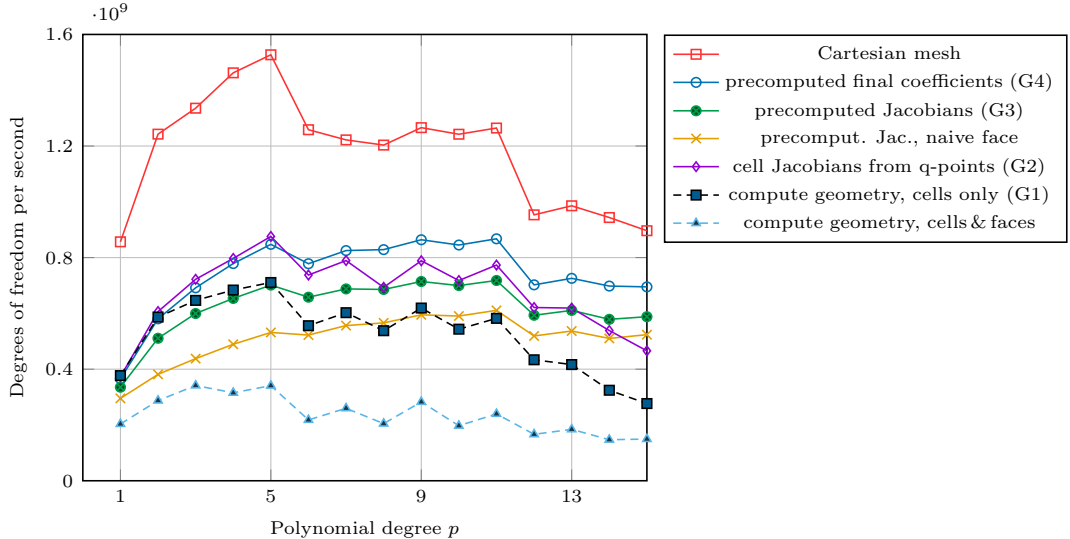
\begin{figure}
  \centering
  \begin{tikzpicture}
    \begin{axis}[
      width=0.6\textwidth,
      height=0.5\textwidth,
      xlabel={Polynomial degree $p$},
      ylabel={Degrees of freedom per second},
      tick label style={font=\scriptsize},
      label style={font=\scriptsize},
      title style={at={(1,0.952)},anchor=north east,draw=black,fill=white,font=\scriptsize},
      legend style={font=\scriptsize},
      legend pos = outer north east,
      ymin=0,ymax=1.6e9,
      xmin=0,xmax=15,
      ytick={0,0.4e9,0.8e9,1.2e9,1.6e9},
      xtick={1,5,9,13},
      grid,
      cycle list name=colorGPL,
      mark size=1.8,
      semithick
      ]
      \pgfplotsset{cycle list shift=1}
      \addplot table[x={degree}, y expr={\thisrowno{1}/\thisrowno{5}}] {\matvecSystems};
      \addlegendentry{Cartesian mesh};
      \pgfplotsset{cycle list shift=-1}
      \addplot table[x={degree}, y={merged}] {\matvecGeometryLaplace};
      \addlegendentry{precomputed final coefficients (G4)};
      \pgfplotsset{cycle list shift=0}
      \addplot table[x={degree}, y={usual}] {\matvecGeometryLaplace};
      \addlegendentry{precomputed Jacobians (G3)};
      \addplot table[x={degree}, y={gradface}] {\matvecGeometryLaplace};
      \addlegendentry{precomput. Jac., naive face};
      \addplot table[x={degree}, y={colloccell}] {\matvecGeometryLaplace};
      \addlegendentry{cell Jacobians from q-points (G2)};
      \addplot table[x={degree}, y={isocell}] {\matvecGeometryLaplace};
      \addlegendentry{compute geometry, cells only (G1)};
      \addplot table[x={degree}, y={isoall}] {\matvecGeometryLaplace};
      \addlegendentry{compute geometry, cells\,\&\,faces};
    \end{axis}
  \end{tikzpicture}
\caption{Throughput of 3D Laplacian kernel for various ways to represent a curved geometry on 28 cores of Broadwell.}
\label{fig:geometry}
\end{figure}

\begin{figure}
\centering
\begin{tikzpicture}
\begin{loglogaxis}[
  grid=both,
  width=0.68\textwidth,
  height=0.52\textwidth,
  xlabel=FLOP/byte ratio,
  ylabel=GFLOP/s,
  axis lines=left,
  ytick={32,64,128,256,512,1024},
  yticklabels={32,64,128,256,512,1024},
  xmin=0.25, xmax=48,ymin=30,ymax=1600,
  xtick={0.25,0.5,1,2,4,8,16,32},
  xticklabels={$\frac 14$,$\frac 12$,1,2,4,8,16,32},
  tick label style={font=\scriptsize},
  label style={font=\scriptsize},
  legend style={font=\scriptsize},
  legend pos = outer north east
  ]
  \addplot[thick, color=black!10, fill=black!25, fill opacity=0.4,forget plot] coordinates{ (0.25,31.25) (10.39,1299) (48,1299)} \closedcycle;
  \draw[draw=black,thick] (axis cs:0.25,31.25) -- node[above,rotate=43.5,inner sep=2pt,outer sep=0.5pt,fill=white]{\tiny Pure load memory bw 125 GB/s}(axis cs:10.39, 1299);
  \draw[draw=black,thick] (axis cs:10.39, 1299) -- (axis cs:48, 1299) node[anchor=south east,inner sep=1pt,outer sep=0.5pt,fill=white]{\tiny Peak DP 2.9 GHz};
  \draw[draw=black,thick] (axis cs:5.20, 650) -- (axis cs:48,650) node[anchor=south east,inner sep=1pt,outer sep=0.5pt,fill=black!10]{\tiny w/o FMA};
  \draw[draw=black,thick] (axis cs:1.30, 162) -- (axis cs:48,162) node[anchor=south east,inner sep=1pt,outer sep=0.5pt,fill=black!10]{\tiny w/o vectorization};
  \addplot[very thin,mark=square,red!80!white,every mark/.append style={semithick}]
    table[x={cartbala},y expr={\thisrowno{0}/500}]{\rooflineLaplaceThreeD}
         [arrow inside={end=stealth,opt={red!80!white, scale=0.8}}{0.2,  0.4, 0.6, 0.85}];
  \addlegendentry{Cartesian mesh};
  \addplot[very thin,mark=otimes*,gnuplot@green,every mark/.append style={semithick,fill=gnuplot@green!80!black}]
    table[x expr={\thisrowno{3}},y expr={\thisrowno{2}/500}]{\rooflineLaplaceThreeD}
         [arrow inside={end=stealth,opt={gnuplot@green, scale=0.8}}{0.25, 0.55, 0.93}];
  \addlegendentry{precomputed Jacobians (G3)};
  \addplot[very thin,mark=x,orange,every mark/.append style={semithick}]
    table[x expr={\thisrowno{5}},y expr={\thisrowno{4}/500}]{\rooflineLaplaceThreeD}
         [arrow inside={end=stealth,opt={orange, scale=0.8}}{0.25, 0.55, 0.95}];
  \addlegendentry{precomp.~Jac., naive face};
  \addplot[very thin,mark=diamond*,gnuplot@purple,every mark/.append style={semithick}]
    table[x expr={\thisrowno{7}},y expr={\thisrowno{6}/500}]{\rooflineLaplaceThreeD}
         [arrow inside={end=stealth,opt={gnuplot@purple, scale=0.8}}{0.25, 0.55, 0.95}];
  \addlegendentry{cell Jacobians from q-pts (G2)};
  \addplot[very thin,gnuplot@lightblue,every mark/.append style={semithick,solid,fill=gnuplot@lightblue!30!black},mark=triangle*]
    table[x expr={\thisrowno{9}},y expr={\thisrowno{8}/500}]{\rooflineLaplaceThreeD}
         [arrow inside={end=stealth,opt={gnuplot@lightblue, scale=0.8}}{0.12, 0.4, 0.8}];
  \addlegendentry{compute geometry, cells\,\&\,faces};
\end{loglogaxis}
\end{tikzpicture}
\caption{Roofline model for the evaluation of the 3D Laplacian with different
  geometry variants on $2\times 14$ cores of Broadwell, displaying the data
  for $p=1, 3, 5, 8, 11$ for five cases from Fig.~\ref{fig:geometry} regarding
  the geometry representation. Higher degrees are located further to the
  right, except for the computation of the full geometry where tensor product
  kernels at $p=11$ spill to main memory at a FLOP/byte ratio of $2.2$. The
  arithmetic balance is based on measured memory throughput and measured
  GFLOP/s rates with the likwid tool. Arrows indicate increasing polynomial degrees.}
\label{fig:roofline}
\end{figure}
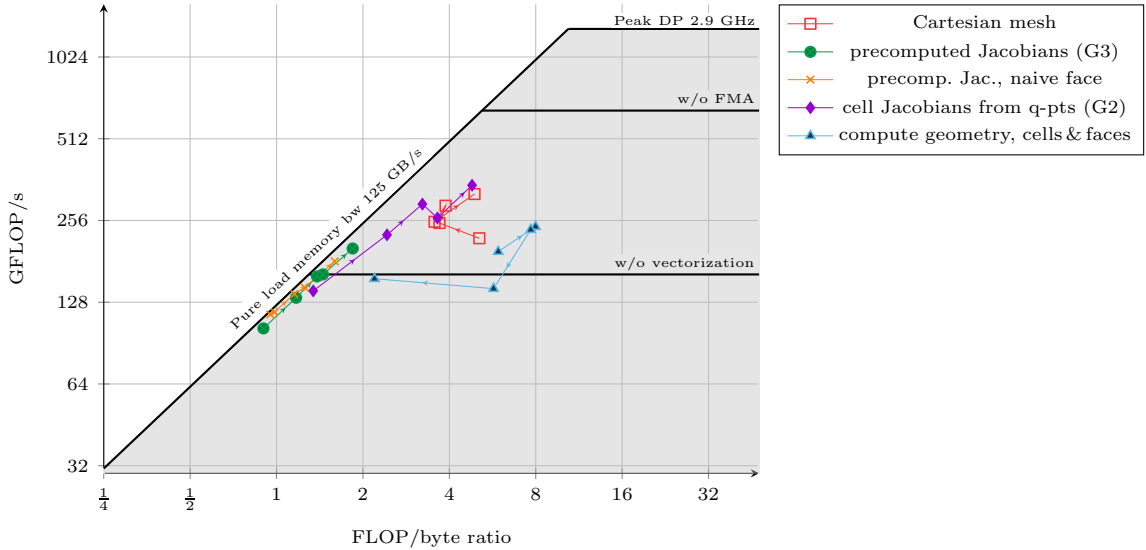

Fig.~\ref{fig:geometry} shows the throughput of the 3D Laplacian with the
various geometry representations listed above on a hypershell mesh and
high-order polynomial geometry descriptions. The problem size is similar to
Cartesian meshes with 8 million to 57 million degrees of freedom. As can be
seen from the performance model in Fig.~\ref{fig:roofline}, the reason for the
gap to the Cartesian mesh case is that the code is clearly memory bandwidth
bound in the variable geometry case. The pre-computed geometry options show
the most consistent performance, despite the highest memory transfer as an
inverse Jacobian and the Jacobian determinant (G3) or final coefficients of
(G4) must be loaded for each quadrature point. A distinctive trend is that the
memory bandwidth limit renders throughput measured as the number of degrees of
freedom processed per second almost constant for a wide range of polynomial
degrees, $3\leq p \leq 11$, since the data accessed per degree of freedom is
$\mathcal O(1)$.

When computing the full geometry of both cells and faces from a continuous FE
space, marked ``compute geometry, cells \& faces'' in Fig.~\ref{fig:geometry}, performance does not exceed 350 million degrees of freedom per
second. Since face integrals only access $k^{d-1}$ data points out of the
$k^d$ elemental degrees of freedom, it is preferable to at least store the
evaluated geometry of the faces as in (G1). Another alternative is provided by the item ``precomputed Jacobian, naive face'' where we first
compute the full gradient
$ \left(\mathcal J^{(e)}\right)^{-\mathrm T} \nabla_{\boldsymbol \xi} u$ and
then multiply by $\ve n$. All variants (G1)--(G4) in Fig.~\ref{fig:geometry} get the
normal gradient $\ve n \cdot \nabla_{\ve x} u$ on faces by computing
$ \ve {j}_{\ve n} \nabla_{\boldsymbol \xi}$ accessing the normal times the
Jacobian
$\ve j_{\ve n} = \ve n \cdot \left(\mathcal J^{(e)}\right)^{-\mathrm T}$
according to Design Choice \ref{req:face_normal_jac}.

\begin{Requirement}\label{req:face_normal_jac}
  For
  discretizations involving the gradient on faces such as the symmetric
  interior penalty DG discretization of the Laplacian \eqref{eq:poisson_sip}
  where the inverse Jacobian $\left(\mathcal J^{(e)}\right)^{-\mathrm T} $ and
  the normal vector $\ve n$ appear together, it is advisable to use access
  functions of the type \texttt{get\_normal\-\_times\_gradient} that can
  directly access a pre-stored vector
  $ \ve {j}_{\ve n} = \ve n \cdot \left(\mathcal J^{(e)}\right)^{-\mathrm T}$ (normal times
  inverse Jacobian). This approach reduces the
  memory access from $2d^2+d$ doubles per quadrature point to $2d$ doubles for
  an interior face.
\end{Requirement}

With respect to the cell
integrals, we see that geometry computation from a continuous finite element
field (G1) that involves gather functions is only competitive for low degrees
where the reduced memory access pays off. For $p>5$, some pre-computation is
indeed preferable. The computation of cell Jacobians from the pre-computed
positions of the quadrature points that only invokes
\texttt{collocation\_derivative} is a very attractive option and even
outperforms the precomputed final coefficients option for low
degrees. However, for large degrees $p>10$ the computations of derivatives
through the collocation approach involves temporary data arrays that go to
outer level caches and eventually spill to main memory.

Fig.~\ref{fig:roofline} displays the measured performance of operator
evaluation in terms of the roofline performance model \cite{williams09}. The pre-computed versions are essentially at the limit of the memory
bandwidth. For the collocation derivative approach that gets the Jacobian from
the quadrature point locations, a considerably higher arithmetic
intensity with some components dominated by memory transfer and others being
compute-bound is observed, in particular at higher degrees. Note that inverting the
Jacobian involves about three times as many multiplications as FMAs.
The Cartesian mesh case is, as discussed in the previous
sections, a bit away from the theoretical performance limits, which is mostly
due to the MPI data exchange that consumes almost one third of the time
at full memory bandwidth limit.

\section{Conclusions and future developments}\label{sec:conclusions}

We have presented a detailed performance analysis of matrix-free operator
evaluation for discontinuous Galerkin methods with cell and face
integrals. The methods are specialized for quadrilateral and hexahedral meshes
and use sum factorization techniques for computing the integrals by
quadrature.  The implementations are
  available to both explicit time integration as well as for the solution of linear and
  nonlinear systems with iterative solvers that spend the bulk of their time in
  matrix-vector products or residual evaluations.
  This work has highlighted the most important algorithmic choices
to reach optimal performance.  We have presented efficient sum factorization kernels
with various compute optimizations that make use of explicit vectorization
over several cells and apply an even-odd decomposition for further reduction of the
arithmetics of the local 1D interpolation. The
resulting local kernels have been shown to reach up to 60\% of arithmetic peak
on Intel Haswell and Broadwell processors and up to 50\% of peak on an Intel
Knights Landing manycore system. For complex geometries, we found that storing
the geometry data on the faces delivers much better throughput than
computation on the fly, whereas computing the inverse Jacobian for the cell
integral from the quadrature point locations by a collocated derivative
algorithm can improve performance over loading precomputed Jacobians from main
memory.

Our experiments show that the proposed highly optimized local kernels render cell integrals alone mostly memory bandwidth
bound for low and moderate polynomial degrees up to ten. Thus, cell and face
integrals must be interleaved for reaching optimal performance. When it comes
to the MPI parallelization, our experiments have identified the MPI data
exchange operations to take up to a third of the operator evaluation time on a single node,
reducing the throughput from around 2 billion degrees of freedom on 28
Broadwell cores (at up to 430 GFLOP/s) to around 1.5 billion degrees of
freedom per second (at up to 320 GFLOP/s), even after optimizing the MPI data
transfer for special polynomial bases. Thus, a future goal is the development
of efficient shared-memory parallelization schemes or MPI shared memory
schemes according to the MPI-3 standard. Furthermore, these alternative
parallelization concepts reduce the amount of duplicated data in general,
promising better use of many-core architectures that have less memory per core available than today's multi-core processors.

{
\setlength{\itemsep}{0pt}
\setlength{\parsep}{0pt}
\setlength{\parskip}{0pt}

}

\end{document}